\documentclass{article}

\usepackage{arxiv}
\usepackage[utf8]{inputenc} 
\usepackage[T1]{fontenc}    
\usepackage{hyperref}       
\usepackage{url}            
\usepackage{booktabs}       
\usepackage{amsfonts}      
\usepackage{nicefrac}       
\usepackage{microtype}      
\usepackage{graphicx}
\usepackage{doi}
\usepackage{lineno}
\usepackage[colorinlistoftodos]{todonotes}
\usepackage{listings}
\usepackage{textcomp}
\usepackage{subcaption}
\usepackage{array}
\usepackage{comment}
\usepackage{mathtools}
\usepackage{calc}
\usepackage{indentfirst}
\usepackage{fancyhdr}
\usepackage{epstopdf}
\usepackage{lastpage}
\usepackage{ifthen}
\usepackage{float}
\usepackage{amsmath}
\usepackage{amssymb}
\usepackage{setspace}
\usepackage{enumitem}
\usepackage{mathpazo}
\usepackage{titlesec}
\usepackage{etoolbox}
\usepackage{tabto}
\usepackage{xcolor}
\usepackage{soul}
\usepackage{multirow}
\usepackage{tikz}
\usepackage{totcount}
\usepackage{changepage} 
\usepackage{paracol}
\usepackage{attrib}
\usepackage{upgreek}
\usepackage{cleveref} 
\usepackage{amsthm}
\usepackage{hyphenat}
\usepackage{footmisc}
\usepackage{geometry}
\usepackage{newfloat}
\usepackage{caption}

\title{Characterisation of SiPM Photon Emission in the Dark}

\date{} 					

\author{J.~B.~McLaughlin$^{1,2,*}$\and G.~Gallina$^{1}$\and F.~Reti\`ere$^{1,3}$\and A.~De~St.~Croix$^{1,6}$\and P.~Giampa$^{1,5}$\and M.~Mahtab$^{1}$\and P.~Margetak$^{1}$\and L.~Martin$^{1}$\and N.~Massacret$^{1}$\and J.~Monroe$^{2}$\and M.~Patel$^{1,3}$\and K.~Raymond$^{1}$\and J.~Roiseux$^{1}$\and L.~Xie$^{1}$\and G.~Zhang$^{4}$}

\renewcommand{\shorttitle}{Characterisation of SiPM Photon Emission in the Dark}

\hypersetup{
pdftitle={Characterisation of SiPM Photon Emission in the Dark},
pdfsubject={physics, instrumentation and detectors},
pdfauthor={J.~B.~McLaughlin, G.~Gallina and F.~Reti\`ere},
pdfkeywords={Silicon Photomultipliers; Multi-Pixel Photon Counters; FBK VUV-HD3; HPK VUV4; Spectroscopy; Microscopy; Dark noise; External cross-talk; nEXO; DarkSide},
}

\begin{document}
\maketitle
\begin{itemize}
\item[\textbf{1}]{TRIUMF, 4004 Wesbrook Mall, Vancouver, BC V6T 2A3, Canada}
\item[\textbf{2}]{Department of Physics, Royal Holloway, University of London, Egham TW20 0EX, UK}
\item[\textbf{3}]{Department of Physics, Simon Fraser University, 8888 University Drive, Burnaby, BC V5A 1S6, Canada}
\item[\textbf{4}]{School of Science, Xi’an Polytechnic University, 710048, China}
\item[\textbf{5}]{SNOLAB, Lively, Ontario, P3Y 1N2, Canada}
\item[\textbf{6}]{Department of Physics, Engineering Physics \& Astronomy, 64 Bader Lane, Queen's University, Kingston, ON, Canada}
\end{itemize}

\bigskip
* joseph.mclaughlin.2018@live.rhul.ac.uk

\begin{abstract}
In this paper, we report on the photon emission of Silicon Photomultipliers (SiPMs) from avalanche pulses generated in dark condition, with the main objective of better understanding the associated systematics for next-generation, large area, SiPM-based physics experiments.  A new apparatus for spectral and imaging analysis was developed at TRIUMF and used to measure the light emitted by the two SiPMs considered as photo-sensor candidates for the nEXO neutrinoless double-beta decay experiment: one Fondazione Bruno Kessler (FBK) VUV-HD Low Field (LF) Low After Pulse (Low AP) (VUV-HD3) SiPM, and one Hamamatsu Photonics K.K. (HPK) VUV4 Multi-Pixel Photon Counter (MPPC). Spectral measurements of their light emission were taken with varying over-voltage in the wavelength range of 450--1020\,nm. For the FBK VUV-HD3, at an over-voltage of $12.1\pm1.0$\,V, we measure a secondary photon yield (number of photons ($\gamma$) emitted per charge carrier ($e^{-}$)) of  $(4.04\pm0.02)\times 10^{-6}$ $\gamma/e^{-}$. The emission spectrum of the FBK VUV-HD3 contains an interference pattern consistent with thin-film interference. Additionally, emission microscopy images (EMMIs) of the FBK VUV-HD3 show a small number of highly localized regions with increased light intensity (hotspots) randomly distributed over the SiPM surface area. For the HPK VUV4 MPPC, at an over-voltage of $10.7\pm1.0$\,V, we measure a secondary photon yield of  $(8.71\pm0.04)\times 10^{-6}$ $\gamma/e^{-}$. In contrast to the FBK VUV-HD3, the emission spectra of the HPK VUV4 don’t show an interference pattern---most likely due to a thinner surface coating. EMMIs of the HPK VUV4 also reveal a larger number of hotspots compared to the FBK VUV-HD3, especially in one of the corners of the device. Note as the photon yield reported in this paper may be limited if compared with the one reported in previous studies due to the measurement wavelength range that is only up to 1020~nm.
\end{abstract}


\section{Introduction}\label{sec:intro}

Silicon photo-multipliers (SiPM)s have emerged as a compelling photo-sensor solution for detecting single photons in applications ranging from particle physics to medical imaging and ranging. SiPMs consist of an array of tightly packaged Single Photon Avalanche Diodes (SPADs) with quenching resistor operated above the breakdown voltage, $V_\mathrm{bd}$, to generate self-sustaining charge avalanches upon absorbing an incident photon. The excess voltage above breakdown is called \textit{over-voltage}, and it is defined as $V_\mathrm{ov} \equiv ( V - V_\mathrm{bd})$, where $V$ is the reverse bias voltage applied to the SiPM. In contrast to the widely used Photomultiplier Tubes (PMTs), SiPMs are low-voltage powered, optimal for operation at cryogenic temperatures, and have low radioactivity \cite{Baudis2018}. Moreover, SiPMs have excellent Photon Detection Efficiency (PDE), not only in the visible and infrared wavelength range, but also for Vacuum Ultra-Violet (VUV) wavelengths \cite{8490731, CAPASSO2020164478}. For these reasons, SiPMs are the baseline solution in the DUNE experiment \cite{FALCONE2021164648}, aiming at precise neutrino oscillation measurements, the DarkSide-20k experiment searching for dark matter \cite{Carnesecchi_2020,aalseth2018darkside}, and the nEXO neutrinoless double-beta decay search experiment \cite{GALLINA2019371}.

The single-photon detection capabilities of SiPMs stems from its extremely high gain, since a single electron-hole pair can generate a charge avalanche on the order of $10^5$--$10^7$ electrons \cite{Acerbi_gain, Piemonte_gain}. An unfortunate byproduct of the avalanche generation process is the emission of secondary photons \cite{PhysRev.100.700}, which in some works on SiPM characterisation are referred to as \textit{cross-talk} photons \cite{SiPMct,NepomukOtte2009}. 

Secondary photons can be correlated with several factors: electric field, impurity concentrations, doping, geometry, etc. \cite{zhang2019light,LinaLiu2020}. Even if an exhaustive list of production mechanisms is not known conclusively at present avalanche emission in silicon seems due to a combination of: (i) indirect interband transitions, (ii) intraband Bremsstrahlung processes and (iii) direct interband
transitions \cite{akil1999multimechanism,gautam1988photon,Lacaita1993}. Each of these mechanisms are responsible for light emission in different spectral regions, i.e. at certain wavelengths. For example avalanche emission below 2 eV seems dominated by indirect interband transitions, between 2~eV and 2.3~eV by intraband Bremsstrahlung and above 2.3~eV by direct interband \cite{akil1999multimechanism,gautam1988photon,Lacaita1993}. The photon energy value for the transition from predominantly indirect interband to predominantly intraband Bremsstrahlung depends on applied electrical field and material properties, while the transition from Bremsstrahlung to direct interband seems occurring at the same photon energy i.e. 2.3~eV \cite{akil1999multimechanism}. 

Secondary photons in SiPMs are responsible for at least three processes: (i) internal cross-talk (ii) external cross-talk and (iii) optically-induced afterpulsing. With internal cross-talk, we refer to secondary photons that subsequently trigger avalanches in neighbouring SPADs of the same SiPM without escaping from the SiPM itself. With external cross-talk we instead refer to secondary photons that escape from the surface of one SPAD and potentially: (i) reflect back into the SiPM at the surface coating interface and trigger avalanches in neighbouring SPADs \cite{gundacker2020silicon}, (ii) transmit through the SiPM surface coating leaving the SiPM.

Finally with optically-induced afterpulsing we refer to secondary photons that trigger avalanches in the same SPAD that originated the primary secondary photon emission during the SPAD recharging time.

Avalanches inside the same SiPM triggered by secondary photons can be simultaneous with the primary one (Direct Cross-Talk (DiCT)) or delayed by several ns (Delayed Cross-Talk (DeCT)) \cite{Boone2017}, and contribute to the total number of correlated avalanches per pulse produced by the SiPM \cite{gallina2021development}. The processes of DiCT and DeCT  are extensively studied in literature with both measurements \cite{Otte2016} and simulation \cite{NepomukOtte2009}. We refer the reader to \cite{Otte2016,Jamil2018,Acerbi2017} for a detailed explanation of the different pulse-counting techniques used to discriminate these processes.

Secondary photon emission outside the SiPM  that originally produced it can be problematic for large surface area, SiPM-based detectors since each SiPM can trigger other SiPMs in their vicinity, thus contributing to the detector background. For this reason it is important to study the SiPM secondary photon emission in order to quantify the systematic effects which hinder the overall detector performance.

This publication aims to study the emission spectra of the secondary photon emission outside the SiPM and its absolute secondary photon yield: number of photons ($\gamma$) emitted per charge carrier ($e^{-}$). Additionally, it investigates the uniformity of the light production over the entire SiPM surface area, identifying regions with heightened light emission intensity (hotspots), in agreement with other studies \cite{engelmann2018spatially}. For this publication we focused on two SiPMs, considered as photo-sensor candidates for the nEXO experiment: one Fondazione Bruno Kessler (FBK) VUVHD Low Field (LF) Low After Pulse (Low AP) SiPM (VUV-HD3), and one Hamamatsu Photonics (HPK) VUV4 Multi-Pixel Photon Counter (MPPC). A complete characterization of the HPK VUV4 and FBK VUV-HD3 are reported in \cite{GALLINA2019371} and \cite{gallina2021development}, respectively. Table\,\ref{tab:SiPM_values} summarizes the SiPMs specification relevant for this work.

\begin{table}[ht]
    \centering
    \caption{Summary of the SiPM specification whose secondary photon emission is studied in this work \cite{CAPASSO2020164478,GALLINA2019371}. The Fill Factor is defined as the ratio between the photon-sensitive area to the total area of the SiPM. The breakdown voltages are extracted from the SiPM I-V curves; defined as the voltage for which the first derivative  with respect to the voltage of the SiPM current (in log space) is at maximum \cite{Nagai2018}.}
    \begin{tabular}{ l c c }
    \hline
        \textbf{Parameter} & \textbf{FBK VUV-HD3} & \textbf{HPK VUV4}\\ \hline
        Total Area & 6$\times$6\,mm$^2$ & 3$\times$3\,mm$^2$ \\
        SiPM Fill Factor & 80\% & 60\% \\
        SPAD pitch & $35\times35\,\mu$m$^2$ & $50\times50\,\mu$m$^2$ \\
        Breakdown Voltage [298\,K] & 31$\pm$1\,V & 52$\pm$1\,V\\
        \hline
    \end{tabular}
    \label{tab:SiPM_values}
\end{table}

The rest of this paper is organized as follows. In Section\,\ref{sec:MandM:miel} we give a brief description of the setup used for measurements of the SiPM secondary photon emission, and we introduce its basic modes of operation. Overviews and analysis of our results for both imaging and spectroscopy of the biased SiPMs are provided in Sections\,\ref{sec:results:imaging}\,\&\,\ref{sec:results:spectroscopy}, respectively. Lastly, in Section\,\ref{sec:conclusion}, we provide some concluding remarks.

\section{TRIUMF Characterization Setup}\label{sec:MandM:miel}
\begin{figure}[ht]
    \centering
    \includegraphics[width=0.55\textwidth]{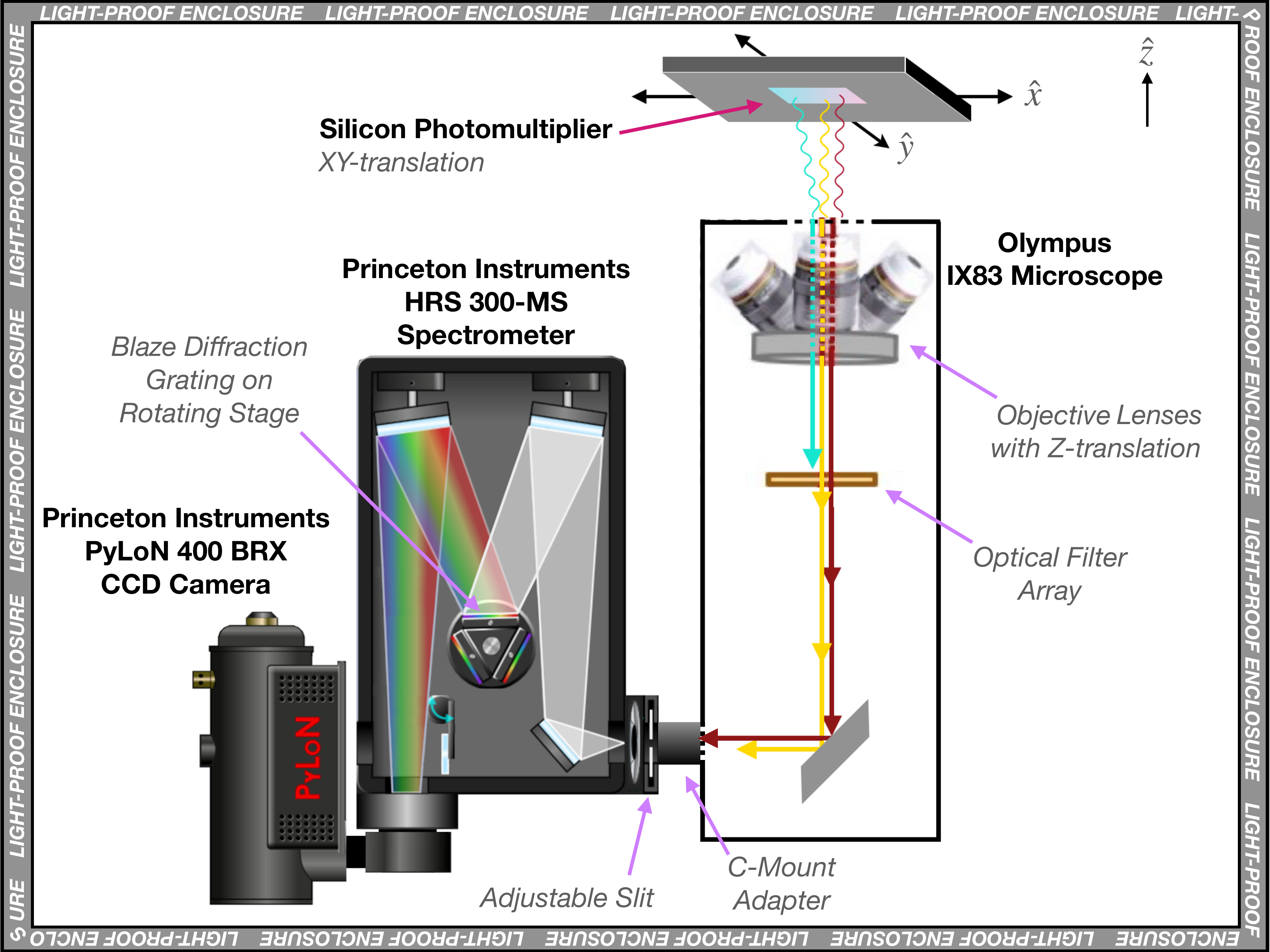}
    \caption{Schematic diagram of the TRIUMF apparatus used for SiPM imaging and spectroscopic measurements. }
    \label{fig:miel_schem}
\end{figure}
A new setup was developed at TRIUMF to characterize the light emitted by SiPMs, as illustrated in Fig.\,\ref{fig:miel_schem}. The setup comprises: (i) an Olympus IX83 microscope, (ii) a Princeton Instruments (PI) HRS 300-MS Spectrometer and (iii) a PI PyLoN\textregistered\, 400BR\_eXcelon CCD camera. The SiPM is affixed to a translation stage above the microscope, with sub-micron motorized position adjustment in the XY-plane of Fig.\,\ref{fig:miel_schem}. The SiPM is biased by a Keithley 6487 Picoammeter, which is also used to monitor the SiPM current overtime. The entire apparatus is contained within a steel, light-proof enclosure, with all the components controlled externally. More precisely, the spectrometer is controlled by the PI LightField\textregistered\ software, while the microscope is controlled by a combination of Olympus software and hardware.

The IX83 microscope incorporates: (i) a filter cube array, used to insert filters in the light path in order to suppress 2$^\mathrm{nd}$- or higher-order diffraction features in the measured spectra (depending on the wavelength range being probed); (ii) an array of objective lenses. Table\,\ref{tab:objectives} summarizes the objective lenses installed in the TRIUMF setup along with their usage, as explained later in this section.

\begin{table}[ht]
    \centering
    \caption{List of Olympus objective lenses used in TRIUMF setup, with their numerical apertures, magnifications, and primary purposes.}
    \begin{tabular}{ l c c c }
        \hline
        \textbf{Lens Model} & \textbf{Magnification} & \textbf{Numerical Aperture} & \textbf{Primary Use} \\ \hline
         PLCN4X-1-7 & 4$\times$ & 0.1 & Imaging \\
         LMPLFLN20X & 20$\times$ & 0.4 & Visible Spectroscopy\\
         LCPLN20XIR & 20$\times$ & 0.45 & NIR Spectroscopy\\ \hline
    \end{tabular}
    \label{tab:objectives}
\end{table}

The PI spectrometer is attached to the microscope via a C-mount adapter and it is equipped with two blaze diffraction gratings: (i) a 300\,lines/mm grating with peak efficiency at 300\,nm and optimal transmission in the near ultraviolet (UV) wavelength range, (ii) a 150\,lines/mm grating with peak efficiency at 800\,nm and optimal transmission in the visible and near-infrared (NIR) wavelength range. These two gratings were chosen to maximize the spectrometer efficiency in the 450--1020\,nm range. Secondary photon emission in the  UV and vacuum ultraviolet (VUV) is in fact expected to be low, as shown in previously reported measurements \cite{mirzoyan2009light}. Additionally, at room temperature, silicon can detect photons only up to 1107.6\,nm due to its band-gap \cite{kittel1996introduction}. Photons with longer wavelengths may still be emitted, as shown in \cite{mirzoyan2009light}, but such photons cannot be detected by SiPMs. Moreover, the PI PyLoN\textregistered\, 400BR\_eXcelon CCD camera that is part of the PI spectrometer system is a silicon-based CCD that is not efficient beyond 1100\,nm. 

\begin{figure}[ht]
    \centering
    \includegraphics[width=0.48\textwidth]{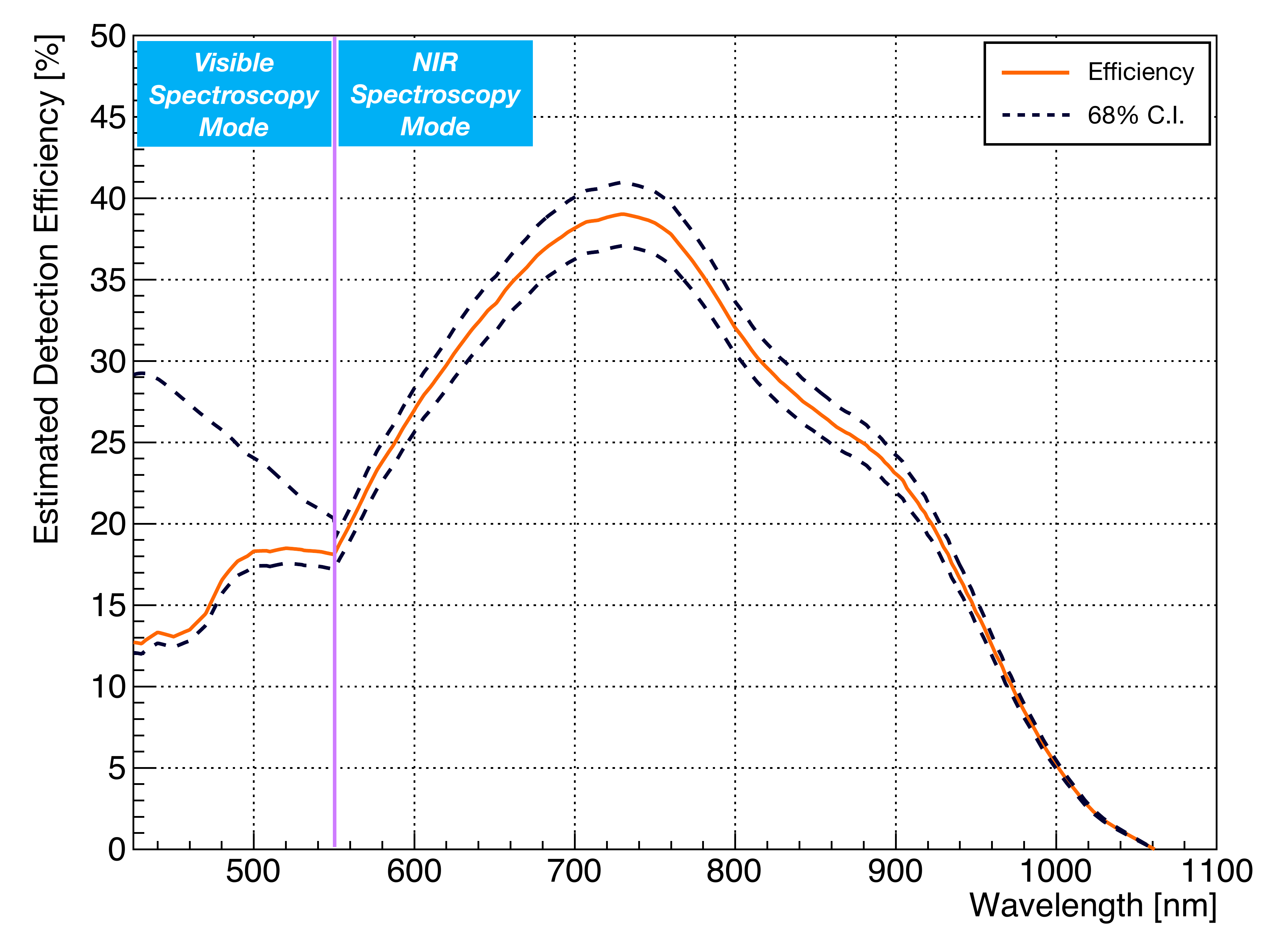}
    \caption{Estimated detection transmission efficiency of the TRIUMF apparatus as a function of the photon wavelength. The 68\% Confidence Interval (C. I.) error bands account for systematic uncertainty in the lamp calibration. Moreover below 550~nm the error increases due to a disagreement between the observed IntelliCal\textregistered\, LED based light source spectrum and its expected spectrum obtained combining the different hardware transmission specifications of the setup.}
    \label{fig:miel_efficiency_full}
\end{figure}

In addition to the camera and gratings, the spectrometer also has an adjustable input slit directly coupled to the C-mount adapter. The slit could also be removed entirely from the optical path, in order to capture emission microscopy images (EMMIs) of the SiPMs. More generally, for the measurements reported in Sec.\,\ref{sec:results:imaging} and Sec.\,\ref{sec:results:spectroscopy} we used the setup in two basic modes of operation, which use different combinations of objective lenses, filters, and gratings, summarized as follows.
\begin{enumerate}
    \item {\textbf{Imaging mode}\\This measurement mode is used to record EMMIs of the biased SiPM in dark conditions, i.e. without external illumination. We used the PLCN4X-1-7 or LMPLFLN20X objectives (depending on the desired field of view) with no optical filters along the light path. Furthermore, the adjustable slit of the PI spectrometer was disengaged and the spectrometer grating (300\,lines/mm) was set to its 0$^\mathrm{th}$-order.\\}
    \item {\textbf{Spectroscopy mode}\\
    The spectroscopy mode is used to measure the spectral components of the secondary photon emitted by the biased SiPM, also under dark conditions. To maximize the transmission of the PI spectrometer and the IX83 microscope in the wavelength range spanning 450--1020\,nm, two combinations of gratings, filters, and microscope objectives were used. Within the 450--550\,nm range, we used the LMPLFLN20X objective lens, no optical filters, and the 300\,lines/mm PI grating centred on its 1$^\mathrm{st}$-order diffraction peak at 500\,nm (this combination is hereafter called Visible Spectroscopy mode). Wavelengths between 550--1020\,nm were measured using the LCPLN20XIR objective, and the 150\,lines/mm grating with its 1$^\mathrm{st}$-order peak centred at 800\,nm. Additionally, a 550\,nm longpass filter was inserted along the IX83 light path to cut any 2$^\mathrm{nd}$-order spectrometer features from wavelengths below 550\,nm (this combination is hereafter called NIR Spectroscopy mode). For the measurements of the SiPM emission spectra reported in Sec.\,\ref{sec:results:spectroscopy}, the slit was set to a width of approximately 200\,$\mu$m, corresponding to a wavelength width in Full Width at Half Maximum (FWHM) of 0.6~nm for the 300\,lines/mm grating, and 1.4~nm for the 150\,lines/mm grating.}
\end{enumerate}
The entire apparatus is calibrated in wavelength and intensity using the PI IntelliCal\textregistered\, calibration system \cite{mcclure2010intellical}. This system includes two light sources:
\begin{itemize}
    \item[(i)]{a Hg and Ne-Ar line source for wavelength calibration with emission lines between 200\,nm and 1000\,nm }
    \item[(ii)]{a NIST traceable LED based light source for relative intensity calibration in the range 450-1020\,nm}
\end{itemize}
To perform these two calibrations, the SiPM was first substituted by the IntelliCal\textregistered\, line source, in order to calibrate the wavelength dependence of the PI spectrometer, and then by the IntelliCal\textregistered\, intensity light source to assess the photon detection efficiency versus wavelength of the entire setup. Fig.\,\ref{fig:miel_efficiency_full} reports 
the measured detection efficiency of the TRIUMF setup as a function of the wavelength.

The error bands account for systematic uncertainty in the lamp calibration. Below 550~nm, the error increases for decreasing wavelength due to a disagreement between the observed IntelliCal\textregistered\,  LED based light source spectrum, measured with the CCD camera, and its expected spectrum obtained combining the different hardware transmission specifications of the setup. The discrepancy is not within the microscope. However, we could not determine if the discrepancy stems from miscalibration of the source or from a significantly lower transmission of the spectrometer+camera system in the visible wavelength range. As the light emission below 550~nm is small \cite{mirzoyan2009light}, the large error band was deemed acceptable.

\section{Imaging of the biased SiPM}\label{sec:results:imaging}
\begin{figure}[ht]
    \centering
    \includegraphics[width=0.73\textwidth]{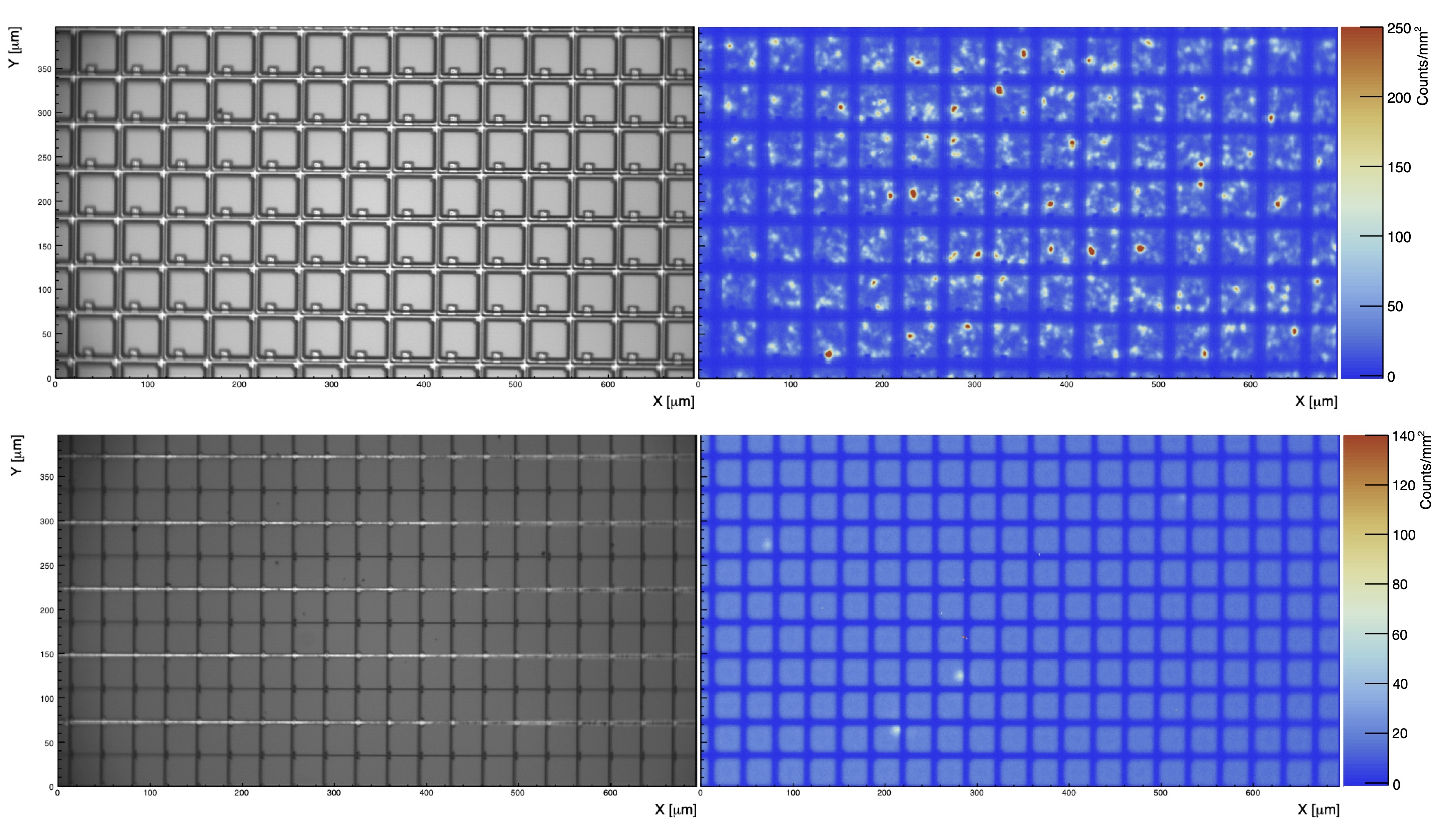}
    \caption{Top row: Emission microscopy image (EMMI) of the HPK VUV4 biased at $11.2\pm 1.0$~V of over voltage. Bottom row: EMMI of FBK 
     VUV-HD3 biased at $13\pm1$~V of over voltage . Both EMMIs were taken with the same camera exposure time and objective lens: LMPLFLN20X (20x magnification).
     }
    \label{fig:LampVsBiased}
\end{figure}

\begin{figure}[ht]
    \centering
    \begin{subfigure}[b]{0.495\textwidth}
         \centering
         \includegraphics[width=\textwidth]{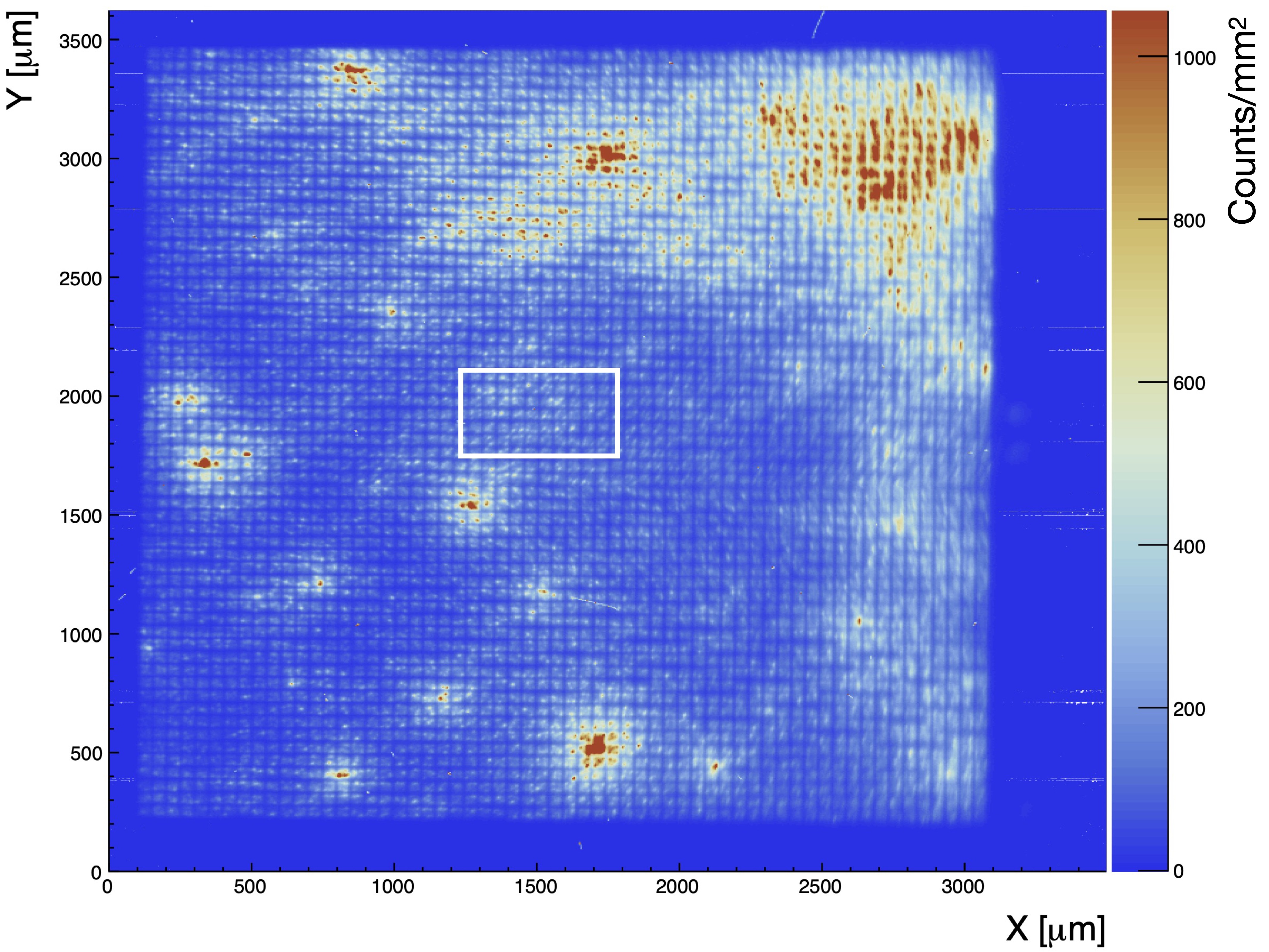}
         \caption{\centering HPK VUV4}
         \label{fig:hpk_fullBias}
     \end{subfigure}
     \hfill
     \begin{subfigure}[b]{0.495\textwidth}
         \centering
         \includegraphics[width=\textwidth]{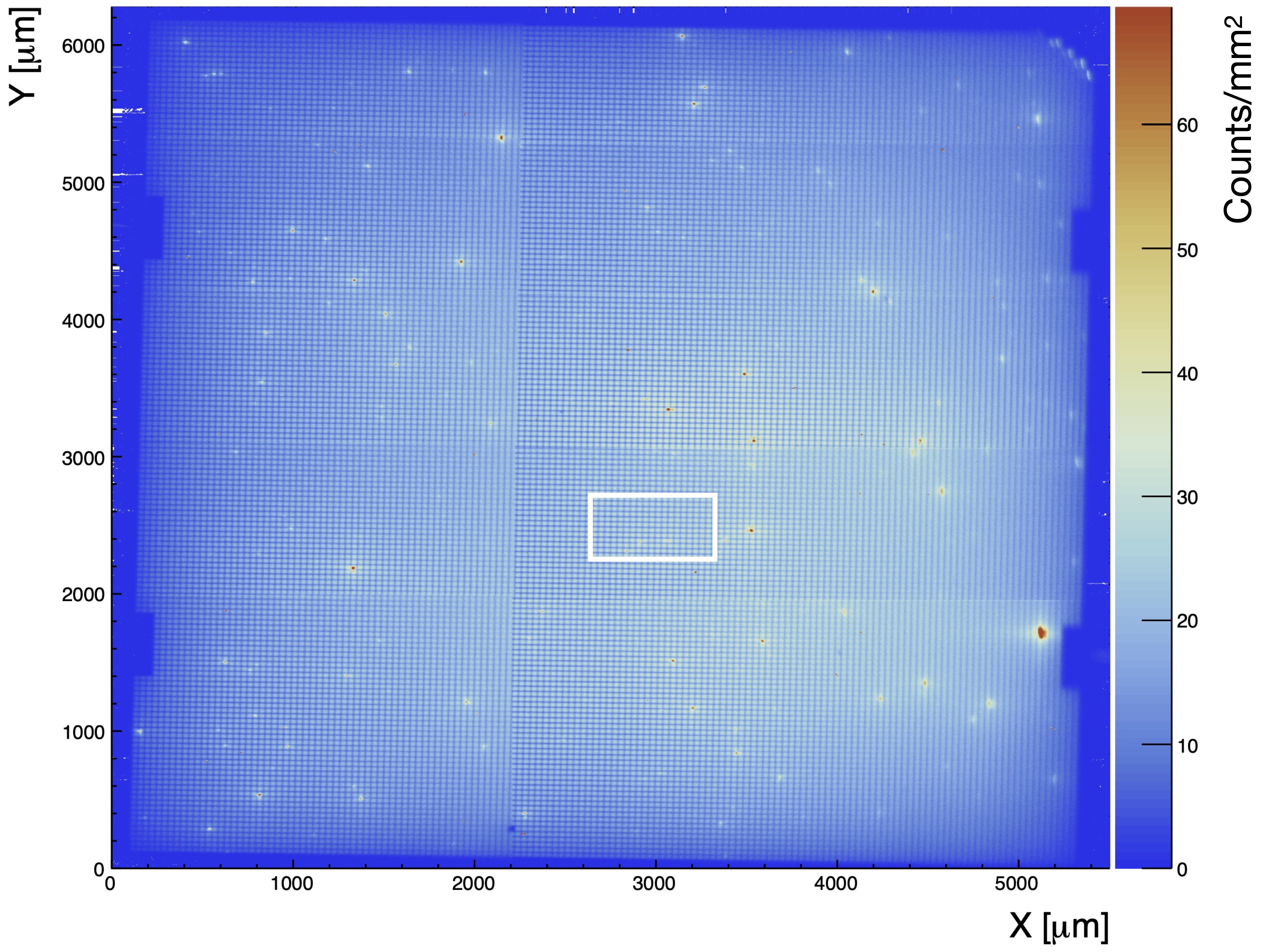}
         \caption{\centering FBK VUV-HD3}
         \label{fig:fbk_fullBias}
     \end{subfigure}
        \caption{Composite EMMI of the HPK VUV4 and FBK VUV-HD3 SiPMs at 4x magnification and $V_\mathrm{ov}= 11.0$\,V and 13.0\,V ($\pm 1.0$\,V), respectively. The regions enclosed in the white boxes are the areas wherein we zoomed to 20$\times$ magnification for spectral measurements. Both EMMI were taken with the same camera exposure time and objective lens: PLCN4X-1-7.}
        \label{fig:fullBias}
\end{figure}

The imaging mode was used to record EMMIs of the biased SiPM as shown in Figs.\,\ref{fig:LampVsBiased}\,\&\,\ref{fig:fullBias}. These EMMIs were used to compare the geometrical fill factors and topographical variations in photon emission for the HPK VUV4 MPPC and the FBK VUV-HD3 SiPM biased at $11.2\pm 1.0$\,V and $13.0\pm1.0$\,V of over-voltage, respectively. At these two over-voltages, the current of the two SiPMs was roughly 2\,mA.  If we make the assumption that the SiPM current is entirely due to charge avalanches\footnote{Leakage current (i.e. not amplified current) could contribute to the total current realized by a SiPM  \cite{Jackson2002}.}, and the photon emission is proportional to the total amount of charge generated by the SiPM, the two SiPM EMMIs in Fig.~\ref{fig:fullBias}, normalised to the same current level per unit area\footnote{The two SiPMs under study have different surface areas therefore Fig.\,\ref{fig:LampVsBiased} and Fig.~\ref{fig:fullBias} were scaled accordingly.}, can be used to compare the relative photon emission intensity and uniformity of the two SiPMs under investigation. 

Since the number of photons emitted per charge carrier is rather low (on the order of $10^{-6}$ $\gamma/e^-$, see Sec.\,\ref{sec:results:spectroscopy}) and this paper focuses on SiPM photon emission driven by avalanche pulses generated in dark conditions, the high over-voltage is needed to generate a sufficiently large number of carriers in the SiPMs such that the light emitted by the SiPM was resulting in a reasonable signal to noise at the PI CCD camera. Fig.\,\ref{fig:fullBias} shows the entire surface of both SiPMs, combining several EMMIs at 4$\times$ magnification. The z-scales in Figs.\,\ref{fig:hpk_fullBias}\,\&\,\ref{fig:fbk_fullBias} show that the HPK VUV4 SiPM tends to have brighter regions with enhanced light intensity (hotspots) compared to the FBK VUV-HD3, for which the hotspots appear more randomly distributed and within single SPADs. More generally the RMS of the light emission of the HPK MPPC is 3.3 times greater than that of the FBK SiPM, and behaves comparably to the one reported in \cite{engelmann2018spatially} for KETEK PM3350T STD/MOD SiPM.  

\section{Spectroscopy of the biased SiPM}\label{sec:results:spectroscopy}
The spectroscopy mode of the TRIUMF setup was used to measure the spectral shape of the secondary photon emitted for each biased SiPM, as introduced in Sec.\,\ref{sec:MandM:miel}. The emission spectra were recorded with the PI LightField\textregistered\, software after calibration of the setup with the IntelliCal\textregistered\,  sources (Sec.\,\ref{sec:MandM:miel}). The net normalization from raw ADC Units (ADU)s recorded by the PI camera, $N_\mathrm{ADU}(\lambda)$, to the number of photons ($\gamma$) emitted per charge carrier, $N_\gamma(\lambda)$ [$\gamma/\text{e}^-$], at a given wavelength ($\lambda$) was obtained as follows

\begin{equation}
    N_\gamma(\lambda) = \frac{N_\mathrm{ADU}(\lambda)\,\eta^\gamma_\mathrm{ADU}\,q_e}{Q_\mathrm{ET}}\bigg(\frac{1}{\rho_\mathrm{surf}\,A(\lambda)\,\epsilon(\lambda)}\bigg),\label{eq:normalization}
\end{equation}

where: $\eta^\gamma_\mathrm{ADU} $ is the calibrated gain of the PI camera equal to $0.7\,\gamma$/ADU \footnote{The camera gain is the independent from the wavelength since it refers to the gain applied by the preamplifier inside the camera after the exposure time has elapsed \cite{PIbook}.}; $q_e$ is the elementary electron charge; $Q_\mathrm{ET}$ is the total charge that passed through the SiPM throughout the fixed exposure time $t_\mathrm{ET}$ defined as 
\begin{equation}
    Q_\mathrm{ET} = \int_0^{t_\mathrm{ET}} i(t)dt
    \label{eq:Qet_integral}
\end{equation}
with $i(t)$ SiPM current\footnote{The SiPM current was monitored during the entire exposure time with the Keithley 6487 Picoammeter (Sec.\,\ref{sec:MandM:miel}).}. The parenthesis in Eq.\,\ref{eq:normalization} represents a correction factor to account for photon losses in the TRIUMF setup. More precisely: $\rho_\mathrm{surf}$ represents the fraction of the SiPM light emitted within the field of view of the spectrometer slit (Sec.\,\ref{sec:results:spectroscopy:Nsurf}), $\epsilon(\lambda)$ is the TRIUMF setup detection transmission efficiency, as shown in Fig.\,\ref{fig:miel_efficiency_full}, and $A(\lambda)$ is a photon acceptance correction factor that accounts for: (i) the finite numerical aperture ($NA$) of the microscope objectives lenses, (ii) reflection and absorption losses due to the SiPM surface coating and to the location of the avalanche region (Sec.\,\ref{sec:results:spectroscopy:I_lambda}).

During acquisition, the camera ADC readout rate was set to 50\,kHz, the lowest speed available, to minimize the ADC readout noise. Table\,\ref{tab:norm_charge} summarizes for each measurement: the exposure time $t_\mathrm{ET}$, the over-voltage $V_\mathrm{ov}$, the average current $\langle i(t)\rangle$ and the total charge $Q_\mathrm{ET}$ that passed through the SiPM throughout the fixed exposure time, as defined by Eq.~\ref{eq:Qet_integral}.
\begin{table}[ht]
    \centering
    \caption{Summary of exposure time ($t_\mathrm{ET}$), over-voltage $V_\mathrm{ov}$, total charge $Q_\mathrm{ET}$ (as defined by Eq.~\ref{eq:Qet_integral}) and average current $\langle i(t)\rangle$ during SiPM spectral measurements.}
    \begin{tabular}{l l c c|c c}
        \textbf{Exposure Time }$t_\mathrm{ET}$& & \multicolumn{2}{c}{\textbf{FBK VUV-HD3}} & \multicolumn{2}{c}{\textbf{HPK VUV4}} \\ \hline
         & & Visible & NIR & Visible & NIR \\ \hline 
         \multirow{3}{*}{8h 20min} & $V_\mathrm{ov}$ [V] & 12.1$\pm$1.0 & 12.1$\pm$1.0 & 10.7$\pm$1.0 & 10.7$\pm$1.0 \\
         & $\langle i(t)\rangle$ [$\mu$A] & 59.1$\pm$1.2 & 66.2$\pm$1.3 & 78.8$\pm$0.6 & 33.3$\pm$0.3 \\
         & $Q_\mathrm{ET}$ [C] & 1.77$\pm$0.12 & 1.99$\pm$0.12 & 2.36$\pm$0.12 & 1.00$\pm$0.12 \\ \hline
         \multirow{3}{*}{4h 45min} & $V_\mathrm{ov}$ [V] & 12.4$\pm$1.0 & 12.4$\pm$1.0 & 10.8$\pm$1.0 & 10.8$\pm$1.0 \\
         & $\langle i(t)\rangle$ [$\mu$A] & 87.1$\pm$0.6 & 97.5$\pm$1.0 & 46.4$\pm$0.4 & 88.0$\pm$0.6 \\
         & $Q_\mathrm{ET}$ [C] & 1.49$\pm$0.09 & 1.67$\pm$0.09 & 0.80$\pm$0.09 & 1.51$\pm$0.09 \\ \hline
         \multirow{3}{*}{3h 20min} & $V_\mathrm{ov}$ [V] & 12.8$\pm$1.0 & 12.8$\pm$1.0 & 11$\pm$1 & 11$\pm$1\\
         & $\langle i(t)\rangle$ [$\mu$A] & 200.9$\pm$1.7 & 165.5$\pm$1.3 & 240.5$\pm$1.3 & 187.8$\pm$1.2 \\
         & $Q_\mathrm{ET}$ [C] & 2.41$\pm$0.08 & 1.99$\pm$0.08 & 2.89$\pm$0.08 & 2.25$\pm$0.08 \\ \hline
    \end{tabular}
    \label{tab:norm_charge}
\end{table}

\subsection{Evaluation of the correction factor $\rho_\mathrm{surf}$}\label{sec:results:spectroscopy:Nsurf}
This section focuses on evaluating the correction factor $\rho_\mathrm{surf}$, used to account for the fraction of the SiPM light emitted within the field of view of the spectrometer slit.
\begin{figure}[ht]
    \centering
    \begin{subfigure}[b]{0.6\textwidth}
        \includegraphics[width=\textwidth]{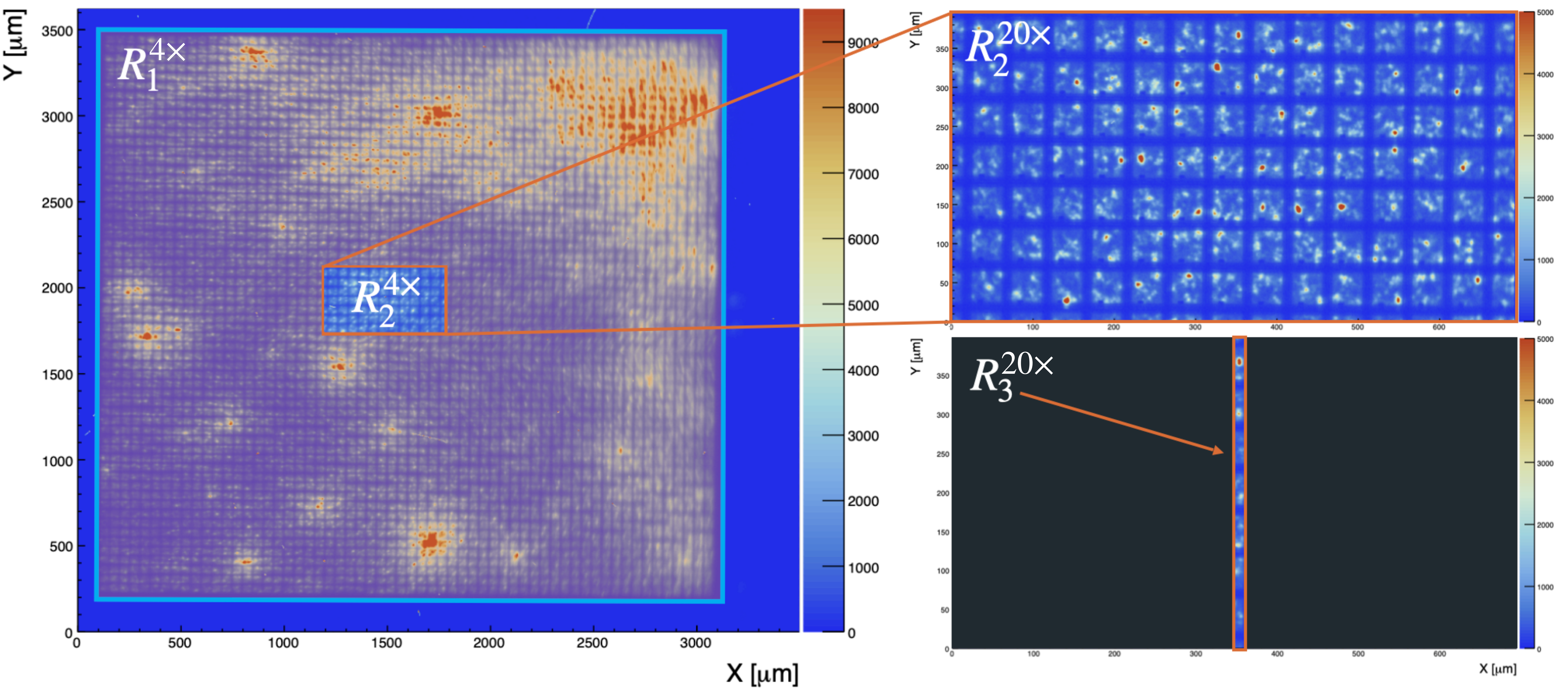}
    \end{subfigure}\hfill
    \begin{subfigure}[b]{0.6\textwidth}
        \includegraphics[width=\textwidth]{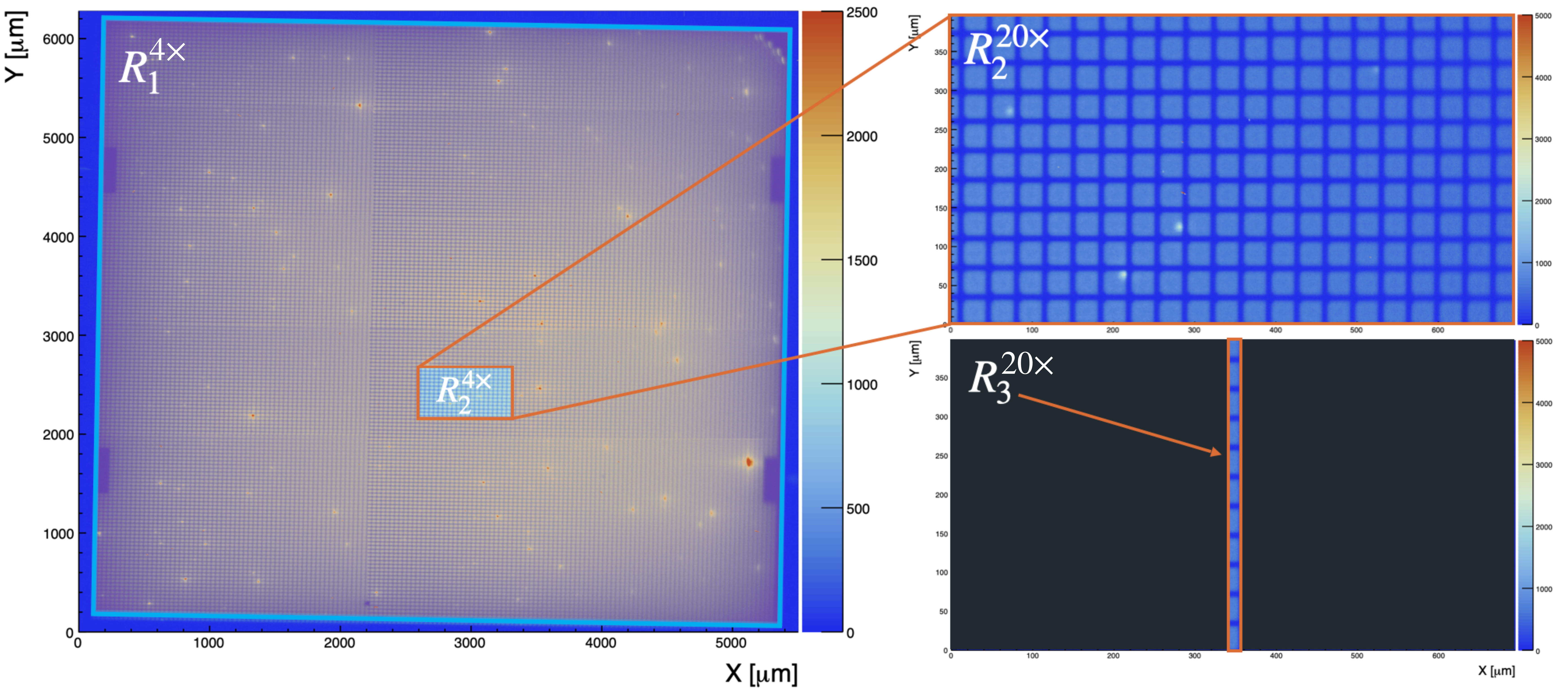}
    \end{subfigure}
    \caption{Pictorial representation of how the quantity $\rho_\mathrm{surf}$ is calculated for the HPK VUV4 MPPC (top) and FBK VUV-HD3 (bottom) SiPM combining the information of Fig.~\ref{fig:LampVsBiased} and Fig.~\ref{fig:fullBias}. $R_k$, with ${k=\{1,2,3\}}$ are selected regions
    where the photon count is measured. See text for more details.
    }
    \label{fig:N_surf}
\end{figure}
$\rho_\mathrm{surf}$ was computed recording EMMIs of the biased SiPM with and without the slit as follows:
\begin{equation}
    \rho_\mathrm{surf} = \bigg(\frac{\sum_{\substack{i,j \subseteq R_3}}P^{20\times}_{i,j}}{\sum_{\substack{i,j \subseteq R_2}} P^{20\times}_{i,j}}\bigg)\bigg(\frac{\sum_{\substack{i,j \subseteq R_2}} P^{4\times}_{i,j}}{\sum_{\substack{i,j \subseteq R_1}} P^{4\times}_{i,j}} \bigg).
    \label{eq:N_surf}
\end{equation}
where: (i) $R_k$, with ${k=\{1,2,3\}}$ are the regions highlighted in Fig.\,\ref{fig:N_surf} for the HPK VUV4 and for the FBK VUV-HD3, and (ii) $P^M_{i,j}$ are the number of photons counted in the $i,j^{th}$ pixel recorded by the CCD camera for an image at magnification $M$. Note as defining $\rho_\mathrm{surf}$ as done in Eq.~\ref{eq:N_surf} not only accounts for the non-uniformity of the light emitted by the SiPM over its entire surface area, but also it removes complications that may arise when comparing EMMIs taken with different magnifications\footnote{The regions $R_2$ shown in Fig.~\ref{fig:N_surf} are the same enclosed in white boxes of Fig.~\ref{fig:fullBias}. These regions are also the ones used in Sec.~\ref{sec:results:spectroscopy} to perform spectral measurements, after insertion of the slit. They were chosen for their centrality and for the absence of bright hotspots. The exact location of these regions is however not relevant since the spectra in Sec.~\ref{sec:results:spectroscopy} are always scaled using Eq.~\ref{eq:N_surf} to account for the non uniformity of the light emitted by the entire SiPM.}. The two ratio of Eq.\,\ref{eq:N_surf} are in fact done between counts of EMMIs taken with the same magnification, i.e. the same objective lens. The $\rho_\mathrm{surf}$ correction factors for the two SiPM tested are reported in Table\,\ref{tab:Nsurf_values}.
\begin{table}[ht]
    \centering
    \caption{Values of $\rho_\mathrm{surf}$ for the FBK VUV-HD3 and HPK VUV4 SiPMs.}
    \begin{tabular}{c|c c}
    \multicolumn{2}{r}{\textbf{FBK VUV-HD3$\quad$}} & {\textbf{HPK VUV4}}\\\hline
    $\rho_\mathrm{surf}$ & $(1.72\pm0.08)\times10^{-4}$ & $(2.40\pm0.12)\times10^{-4}$\\\hline 
    \end{tabular}
    \label{tab:Nsurf_values}
\end{table}
\subsection{Evaluation of the correction factor $A(\lambda)$}\label{sec:results:spectroscopy:I_lambda}

\begin{figure}[ht]
    \centering
    \begin{subfigure}[b]{0.38\textwidth}
        \includegraphics[width=\textwidth]{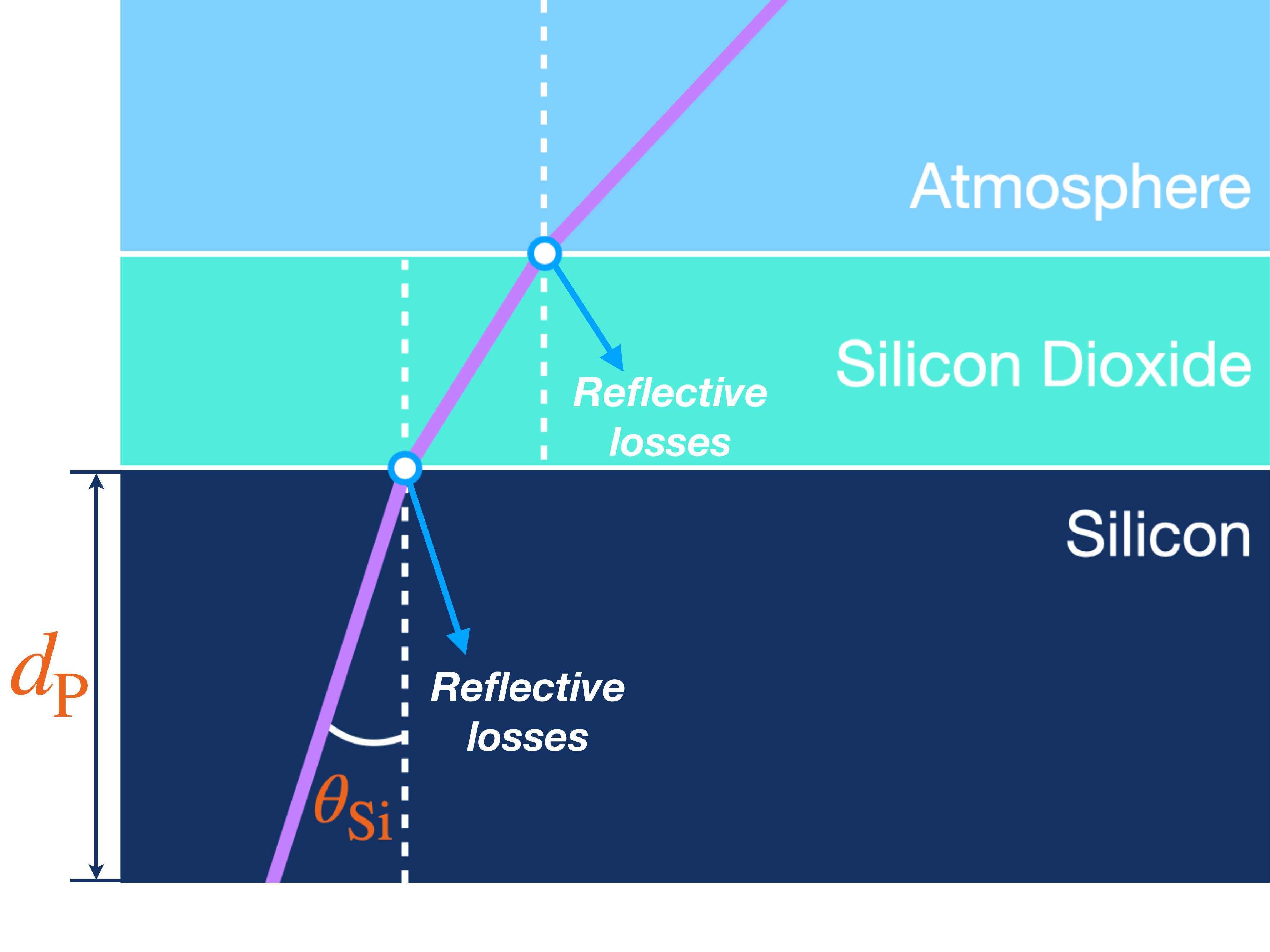}
       \caption{\centering}
        \label{fig:photon_acceptance:snell}
    \end{subfigure}
    \hfill
    \begin{subfigure}[b]{0.52\textwidth}
        \includegraphics[width=\textwidth]{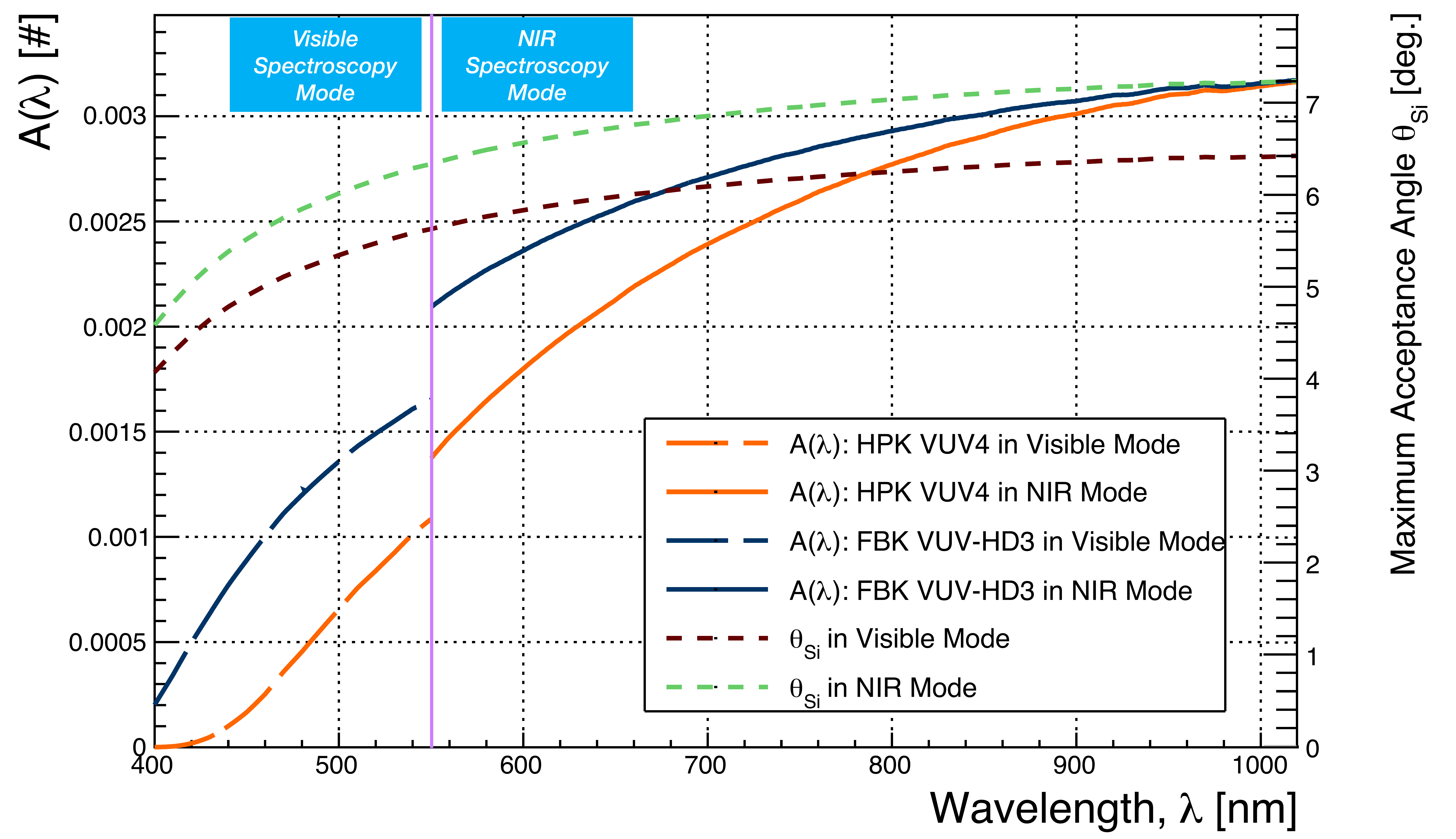}
        \caption{\centering}
        \label{fig:photon_acceptance:Fcorr_lambda}
    \end{subfigure}
    \caption{a) Schematic representation of the SiPM surface coating structure used to compute the photon acceptance correction factor $A(\lambda)$.
    $\theta_\mathrm{Si}$ is the maximum angle for which photons emitted in the silicon can be detected by the microscope objective. $d_\mathrm{P}$ is the depth of the avalanche region. b) Photon acceptance correction factor $A(\lambda)$ (Eq.~\ref{eq:I_lambda}) and
   maximum acceptance angle ($\theta_\mathrm{Si}$, Eq.\,\ref{eq:theta_max}) as a function of the wavelength for the two spectroscopy modes introduced in Sec.\,\ref{sec:MandM:miel}. The discontinuity in the $ A(\lambda)$ correction factor for the two SiPMs is due to the two different objective lenses (with different numerical aperture) used in the Visible (LMPLFLN20X) and NIR (LCPLN20XIR) spectroscopy measurement modes as shown in Sec.\,\ref{sec:MandM:miel}.}
    \label{fig:photon_acceptance}
\end{figure}

The photon acceptance correction factor $A(\lambda)$ accounts for: (i) the finite numerical aperture ($NA$) of the microscope objectives lenses, (ii) reflection and absorption losses due to the SiPM surface coating and to the location of the avalanche region. In what follows we assume that the light emitted by SiPM avalanches is isotropic and not polarized. Additionally, the correction factor $A(\lambda)$ is computed considering a SiPM surface coating structure constituted by a single layer of SiO$_2$, as shown in Fig.~\ref{fig:photon_acceptance:snell}. This structure was provided by FBK and used in \cite{VUVHD3_reflectance} for a study of the SiPM reflectivity. Hamamatsu didn't disclose the HPK VUV4 surface coating structure, and therefore due to lack of more detailed information, we assume the HPK MPPC has the same coating structure as the FBK VUV-HD3 SiPM. The correction factor $A(\lambda)$ was then computed neglecting interference and integrating over the solid angle \cite{griffiths1962introduction} contained within the numerical aperture ($NA$) of the objective lens of the microscope as follows

\begin{align}
    A(\lambda) &= \frac{\int_0^{2\pi}d\phi}{\int_0^{2\pi}d\phi\int_0^{\pi}\sin\theta d\theta}\int_0^{\theta_\mathrm{Si}} e^{-\frac{d_\mathrm{P}}{\cos\theta\,\mu(\lambda)}}(1 - R_\mathrm{Si,SiO_2}(\lambda,\theta))(1 - R_\mathrm{SiO_2,Atm}(\lambda,\theta'))\sin\theta d\theta\\
    & = \frac{1}{2}\int_0^{\theta_\mathrm{Si}} e^{-\frac{d_\mathrm{P}}{\cos\theta\,\mu(\lambda)}}(1 - R_\mathrm{Si,SiO_2}(\lambda,\theta))(1 - R_\mathrm{SiO_2,Atm}(\lambda,\theta'))\sin\theta d\theta,
    \label{eq:I_lambda}
\end{align}
where: (i) $e^{-\frac{d_\mathrm{P}}{\cos\theta\,\mu(\lambda)}}$ (with $\mu(\lambda)$ attenuation length) is a correction factor to account for the self-absorption of the emitted photons in the silicon within a length  $d_\mathrm{P}$, as shown in Fig.~\ref{fig:photon_acceptance:snell}. \footnote{The avalanche region of each SiPM SPAD is located at a certain depth ($d_\mathrm{P}$) from the SPAD surface and the emitted photons need to travel a length equal to this depth before to reach the surface and escape from it. This self-absorption mechanism is significant for wavelength below 450~nm due to the short attenuation lengths of UV photons in silicon \cite{green_self-consistent_2008}, but it is negligible for longer wavelengths. The exact location of the avalanche region was not provided by FBK and HPK, however with the model developed in \cite{8818357} we can infer a lower limit to its depth. We will use therefore the two depths reported in \cite{8818357} to estimate $d_\mathrm{P}$. More precisely for the HPK VUV4 we used  $d_\mathrm{P}=0.8\pm0.2\,\mu$m, while for the FBK VUV-HD3 (that shares with the FBK VUV-HD1 studied in \cite{8818357} the same surface coating and cell structure) $d_\mathrm{P}=0.145\pm0.01\,\mu$m. The wavelength dependent attenuation length was computed accordingly to the data reported in \cite{green_self-consistent_2008}.} (ii) $\phi$ is the azimuthal angle, (iii) $R_\mathrm{Si,SiO_2}$ is the reflectance at the silicon ($\mathrm{Si}$)-silicon dioxide ($\mathrm{SiO_2}$) interface, (iv) $R_\mathrm{SiO_2,Atm}$ is the reflectance at the silicon dioxide ($\mathrm{SiO_2}$)-atmosphere ($\mathrm{Atm}$) interface. Both quantities were computed as reported in \cite{griffiths1962introduction}. $\theta$ is the emission angle of photons in the Silicon, and $\theta_\mathrm{Si}$ is the maximum angle for which photons emitted in the silicon can be detected by the microscope objective. This last quantity is reported in Fig.\,\ref{fig:photon_acceptance:Fcorr_lambda} and it is determined using the definition of $NA$ and the Snell's law \cite{griffiths1962introduction} as follows
\begin{align}
    NA &\equiv n_\mathrm{SiO_2}\,\sin\theta_\mathrm{SiO_2}= n_\mathrm{Si}\,\sin\theta_\mathrm{Si}\label{eq:numerical_aperture}\\
    \therefore\quad\theta_\mathrm{Si}&=\sin^{-1}\bigg(\frac{NA}{n_\mathrm{Si}}\bigg).\label{eq:theta_max}
\end{align}

with: $n_i$ and $\theta_i$ ($i=\{\mathrm{Si},{\mathrm{SiO_2}}\}$) being the refractive indices and photon angles (measured from the normal of the layer boundaries) of the Silicon ($\mathrm{Si}$) and Silicon dioxide ($\mathrm{SiO_2}$) medium. $\theta'$ is similarly defined using Snell’s law as follows,
\begin{equation}
    \theta' = \sin^{-1}\bigg(\frac{n_\mathrm{Si}(\lambda)}{n_\mathrm{SiO_2}(\lambda)} \sin(\theta)\bigg).
    \label{eq:theta_prime}
\end{equation}

Eq.\,\ref{eq:I_lambda} is solved numerically using the refractive index data for each wavelength reported in \cite{aspnes1983dielectric,rodriguez2016self}. Fig.\,\ref{fig:photon_acceptance:Fcorr_lambda} shows the correction factor $ A(\lambda)$ as a function of the wavelength.  The discontinuity in the $ A(\lambda)$ correction factor for the two SiPMs is due to the two different objective lenses (with different numerical aperture) used in the Visible (LMPLFLN20X) and NIR (LCPLN20XIR) spectroscopy measurement modes, as shown in Sec.\,\ref{sec:MandM:miel}. Moreover the NIR spectroscopy mode has a higher $ A(\lambda)$ i.e. smaller correction factor, due to the higher objective lens $NA$.

\subsection{Evaluation of the SiPM photon yields}\label{sec:results:spectroscopy:gain}
\begin{figure}[ht]
    \centering
    \begin{subfigure}{0.475\textwidth}
        \includegraphics[width=\textwidth]{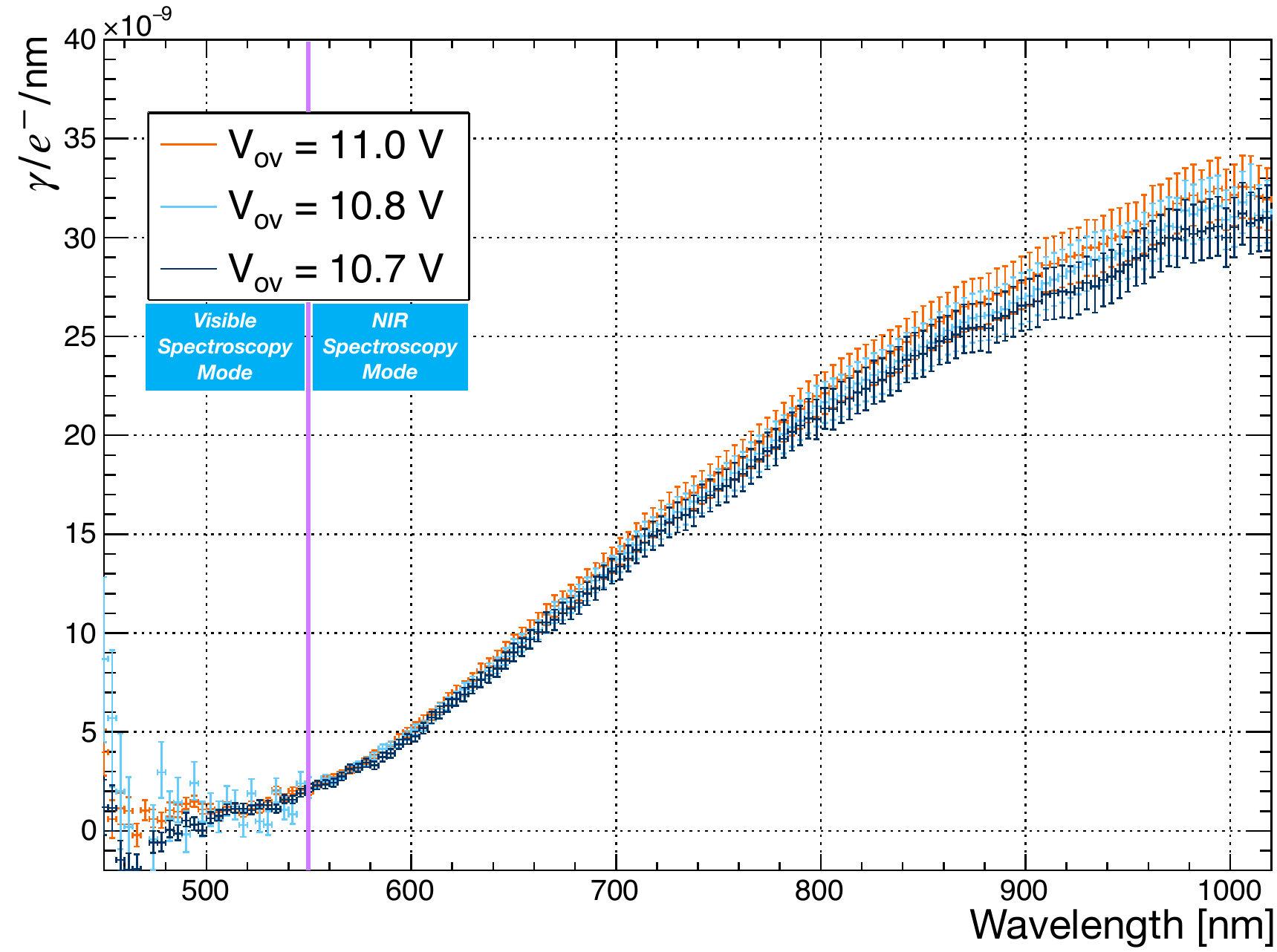}
        \caption{\centering HPK VUV4}
        \label{fig:HPKVUV4_sameSiPM}
    \end{subfigure}\hfill
    \begin{subfigure}{0.475\textwidth}
        \includegraphics[width=\textwidth]{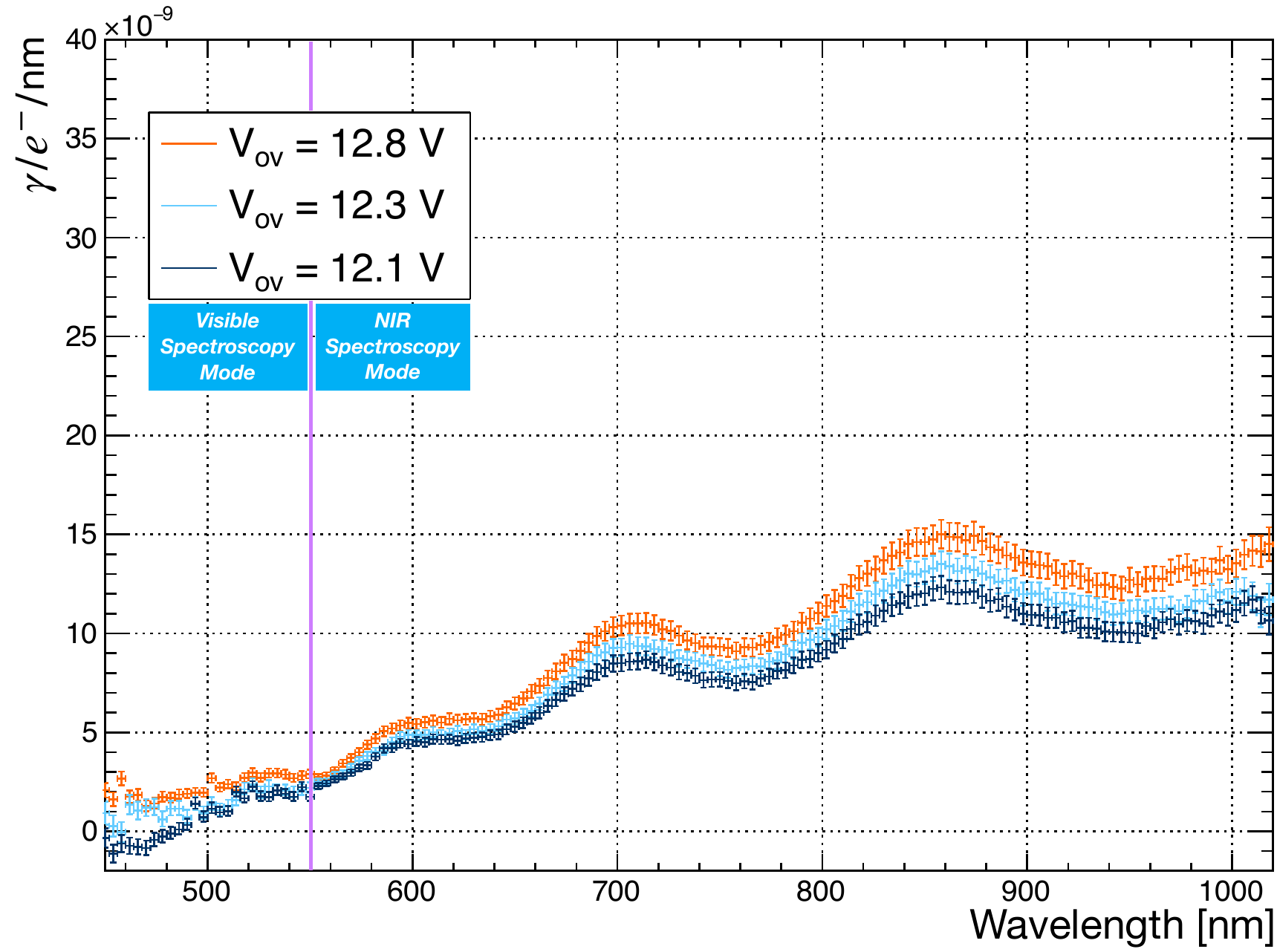}
        \caption{\centering FBK VUV-HD3}
        \label{fig:FBKVUV-HD3_sameSiPM}
    \end{subfigure}
    \caption{Spectra of the HPK VUV4 and FBK VUV-HD3 SiPMs as a function of applied over-voltage.}
    \label{fig:spectra_sameSiPM}
\end{figure}

Figs.\,\ref{fig:HPKVUV4_sameSiPM}\,\&\,\ref{fig:FBKVUV-HD3_sameSiPM} report, for the two SiPM tested, the number of secondary photons emitted per charge carrier per nm $N^{*}_\gamma(\lambda)$, defined as
\begin{equation}
N^{*}_\gamma(\lambda)= \frac{N_\gamma(\lambda)}{\Delta\lambda},
\end{equation}
where $\Delta\lambda$ represents the wavelength resolution, equal to 4\,nm.
The uncertainties in Fig.\,\ref{fig:spectra_sameSiPM} were computed using a Monte Carlo simulation, assuming that: (i) the photon emission of the SiPM follows a Poisson distribution, (ii) the systematic error of the efficiency correction is normally distributed, and (iii) the detection probability of each CCD pixel of the PI camera follows a binomial distribution\footnote{Note, as shown in Sec.\,\ref{sec:MandM:miel}, as the measured wavelength range was studied with two basic modes of operation that comprise different sets of objective lenses, filters, and gratings to maximize the setup detection efficiency. These two modes are defined as Visible Spectroscopy ([450-550]nm) and NIR Spectroscopy ([550-1020]~nm), depending on the wavelength range being studied (Sec.\,\ref{sec:MandM:miel}). After correction for the $A(\lambda)$ factor, residual discontinuities in the spectra of Fig.~\ref{fig:spectra_sameSiPM} were removed by measuring with the PI camera the wavelength range [550-640]~nm with both the Visible and NIR Spectroscopy modes and averaging the CCD counts in order to obtain a smooth transition between the two modes.}

The spectra in Fig.\,\ref{fig:spectra_sameSiPM} show that the secondary photon emission is: (i) predominantly in the red and NIR, (ii) it has a cutoff between 450\,nm and 500\,nm, (iii) it increases with increasing wavelength, and over-voltage. Since $N^{*}_\gamma(\lambda)$ is independent of the number of charge carriers flowing through the SiPM and the SiPM electric field increases with increasing over-voltage, the higher $N^{*}_\gamma(\lambda)$ could be related to electric field dependent processes that contribute to the overall light production, as shown in \cite{zhang2019light,gautam1988photon,wolff1960theory,haecker1974infrared,figielski1962proc}. The increasing $N^{*}_\gamma(\lambda)$ for increasing wavelength agrees with previously reported studies \cite{mirzoyan2009light,akil1999multimechanism,gautam1988photon,Lacaita1993}. The FBK VUV-HD3, in particular, shows a clear signature of oscillations that are due to thin-film interference of light traversing the SiO$_2$ surface coating. The HPK VUV4 doesn't show instead an interference pattern, probably due to its thinner coating. Geometry and surface coating can therefore contribute significantly to the final spectral shape. 

The presence (FBK VUV-HD3) and absence (HPK VUV4) of an interference pattern was also  measured in \cite{VUVHD3_reflectance} during reflectivity measurements for the HPK VUV4 MPPC and the previous generation of FBK SiPMs: the FBK VUV-HD1, which shares with the FBK VUV-HD3 the same surface coating and cell structure. 

The spectra in Fig.\,\ref{fig:spectra_sameSiPM} were measured up to 1020~nm. This is a consequence of the low Estimated Detection Efficiency (EDE) of the TRIUMF setup above 1020~nm, as shown in Fig.~\ref{fig:miel_efficiency_full}. The EDE of the TRIUMF setup is in fact $\sim2.6\%$ at 1020~nm and $<1\%$ for wavelength above 1050~nm.\footnote{Considering a 200\,$\mu$m spectrometer slit (Sec.~\ref{sec:results:spectroscopy}) the signal to noise (S/N) for the longest exposure time (i.e. lowest over voltage) at the PI CCD camera was at 800~nm  between 6 and 8 depending on the SiPM under study, and 1 at 1020~nm. Increasing further the slit would have increased the S/N but would have not improved significantly the capability of the TRIUMF setup to measure spectra above 1020~nm since the estimated detection efficiency of the system is low above 1020~nm.}

Fig.\,\ref{fig:spectra_sameSiPM} can additionally be used to compute the total number of secondary photons emitted per charge carrier (i.e. the SiPM secondary photon yield) by integrating $N^{*}_\gamma(\lambda)$ over the measured emission spectrum as follows 
\begin{equation}
\label{eq:ngamma}
    N_\gamma = \int_{450\text{nm}}^{1020\text{nm}} N^{*}_\gamma(\lambda)\,d\lambda
\end{equation}
Results are reported in Table\,\ref{tab:photon_yields} and compared with the ones reported in \cite{mirzoyan2009light} and \cite{Lacaita1993} measured using a S10362-11-100U HPK MPPC  and a photo-diode, respectively. 

A quantitative comparison with the results reported in \cite{mirzoyan2009light} is not possible since the author didn't provide information on the average current that was flowing in the SiPM during their measurement. The $N_\gamma$ reported in this work is instead smaller if compared with the one reported in \cite{Lacaita1993}, that cover a similar reverse current range (Table\,\ref{tab:norm_charge}). A possible explanation could be searched in the different spectral range covered by the two studies. In \cite{Lacaita1993} the authors in fact measured up to 1087\,nm while we limited our analysis up to 1020~nm, due to the limited efficiency of the TRIUMF setup above this wavelength value (Fig.~\ref{fig:miel_efficiency_full}). Accordingly to \cite{Lacaita1993} the $N_\gamma$ keep increasing for increasing wavelength and therefore limiting the integral of Eq.~\ref{eq:ngamma} up to 1020~nm could have affect the estimation of the SiPM secondary photon yield resulting in a lower value if compared with the one reported in \cite{Lacaita1993}.

\begin{figure}[ht]
    \centering
    \includegraphics[width=0.48\textwidth]{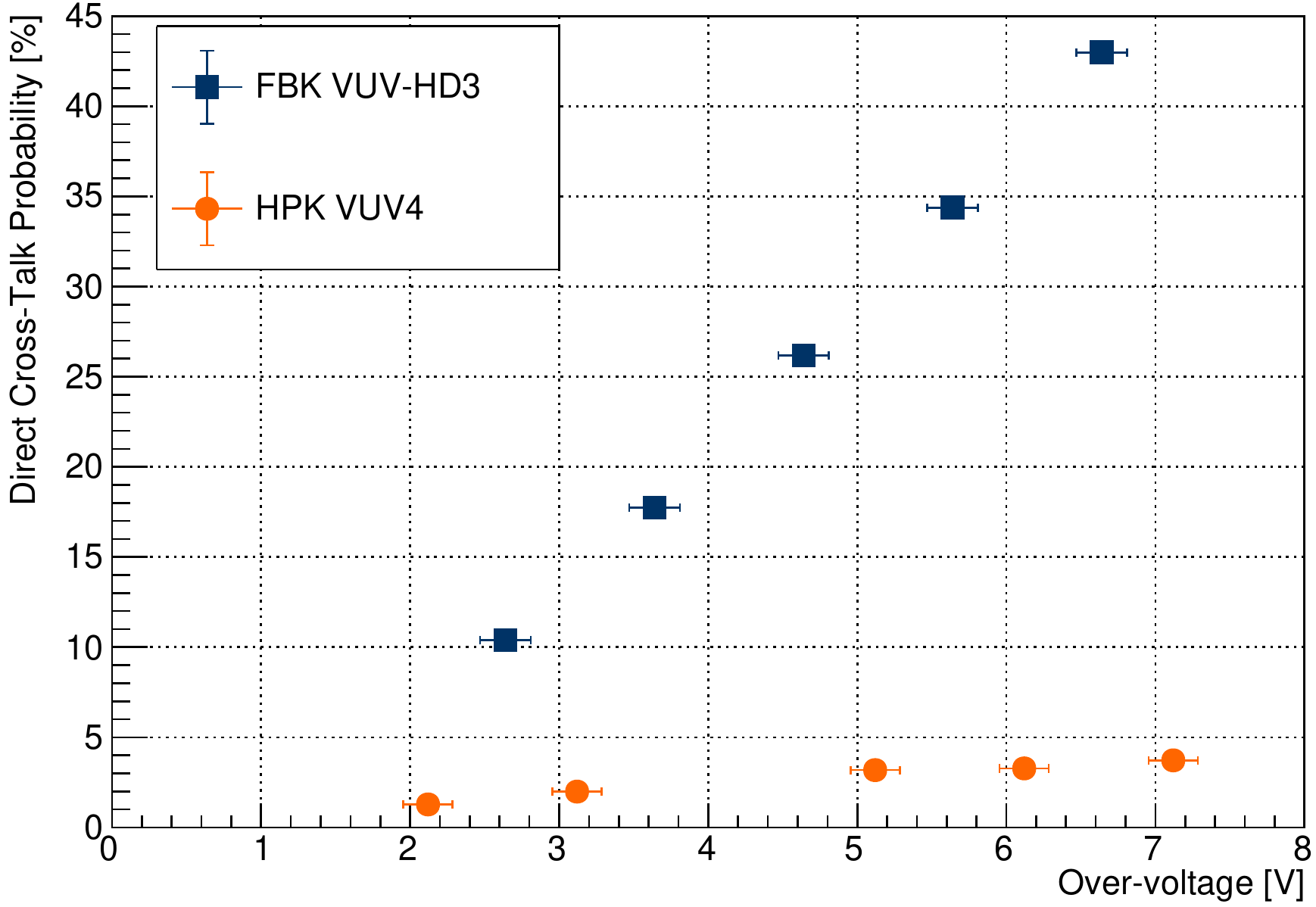}
    \caption{Direct cross-talk probability (DiCT) as a function of the applied over-voltage
    measured at 163\,K for the two SiPMs tested in this work \cite{GALLINA2019371,gallina2021development}.
    }
    \label{fig:DiCT_prob}
\end{figure}

\begin{table}[ht]
    \centering
    \caption{Photon yields (number of photons emitted per charge carrier) measured in the wavelength range [450-1020]~nm for the FBK VUV-HD3 and HPK VUV4 SiPMs as a function of the applied over-voltage. The last two line of the table represent the photon yields measured in \cite{mirzoyan2009light} and \cite{Lacaita1993}, respectively.}
    \begin{tabular}{c c|c c}
         \multicolumn{2}{c}{\textbf{FBK VUV-HD3}}& \multicolumn{2}{c}{\textbf{HPK VUV4}}\\
         \hline 
        $V_\mathrm{ov}$ [V] & Photon Yield [$\gamma/e^{-}$] & $V_\mathrm{ov}$ [V] & Photon Yield [$\gamma/e^{-}$]\\ \hline
        12.1$\pm$1.0 & $(4.04\pm0.02)\times 10^{-6}$ & 10.7$\pm$1.0 & $(8.71\pm0.04)\times 10^{-6}$\\
        12.4$\pm$1.0 & $(4.45\pm0.02)\times 10^{-6}$ & 10.8$\pm$1.0 & $(8.98\pm0.06)\times 10^{-6}$\\
        12.8$\pm$1.0 & $(5.10\pm0.02)\times 10^{-6}$ & 11.0$\pm$1.0 & $(9.24\pm0.05)\times 10^{-6}$\\\hline
        \multicolumn{4}{l}{Photon Yield in \cite{mirzoyan2009light} (500--1117\,nm):$\quad1.2\times10^{-5}$\,$\gamma/e^{-}$}\\\hline
         \multicolumn{4}{l}{Photon Yield in \cite{Lacaita1993} [0.5-4.5] mA (413--1087\,nm):$\quad2.9\times10^{-5}$\,$\gamma/e^{-}$}\\\hline
    \end{tabular}
    \label{tab:photon_yields}
\end{table}
Overall the  number of secondary photons emitted per charge carrier by the HPK VUV4 is roughly a factor of two greater than that of the FBK VUV-HD3 SiPM. However the same is not true for internal cross-talk (i.e. DiCT, Sec.\,\ref{sec:intro}) since, as reported in Fig.\,\ref{fig:DiCT_prob} the DiCT probability of the HPK VUV4 MPPCs is of the order of 3\% at $V_\mathrm{ov} = 4$\,V, while the DiCT probability of the FBK VUV-HD3 is around 20\% for the same $V_\mathrm{ov}$\footnote{The DiCT probability is measured as the ratio between the number of prompt (or trigger) pulses with an integrated charge bigger than 1.5 Photo-electron Equivalent (PE) divided by the number of prompt pulses with an integrated charge bigger than 0.5 PE.}. From this last point, we can deduce that: (i) HPK trenches are extremely effective in suppressing internal cross-talk relative to FBK trenches \cite{Piemonte2016}; (ii) the reduction of the SiPM secondary photon emission doesn't necessarily follow the same design optimization loop compatible with the reduction of DiCT.

Note as the DiCT probabilities reported in Fig.\,\ref{fig:DiCT_prob} were measured at 163\,K since the HPK VUV4 and FBK VUV-HD3 were tested in pulse counting mode in the context of the nEXO experiment. 

The data reported in Fig.\,\ref{fig:spectra_sameSiPM} (and the results of Table\,\ref{tab:photon_yields}) can be used as sampling distributions for a Monte Carlo simulation to estimate the probability of photon emission at a given wavelength per avalanche by each SiPM in the detector. Furthermore, paired with careful measurements of the SiPM Photon Detection Efficiency (PDE) \cite{8818357} in the IR and NIR, this figure contains enough information to estimate the contribution of the SiPM secondary photon emission on the total background rate for any large-area SiPM-based detectors; this is crucial for experiments like nEXO and DarkSide-20k, where the SiPMs will likely be arranged facing each other.

We conclude this section stressing that, in this publication, we studied the SiPM secondary photon emission solely from dark noise-induced avalanches. The high over-voltage was then needed to ensure a reasonable signal to noise at the PI CCD camera. The number of secondary photons per charge carrier per nm reported in Fig.\,\ref{fig:spectra_sameSiPM} could therefore differ from the one emitted at lower over voltages since, on average, SiPMs in pulse counting mode are operated at much smaller over-voltages than the ones reported in Table\,\ref{tab:photon_yields}. The data in Fig.\,\ref{fig:spectra_sameSiPM} are however normalized to the total generated charge in the SiPM and therefore, in principle, independent of the SiPM gain. Future analysis will focus on the SiPM secondary photon emission induced by laser-driven avalanches. This will allow not only to study light propagation between neighbouring SPADs in the SiPMs, but also to probe the emission spectrum at lower over-voltages.
\section{Conclusion}\label{sec:conclusion}

SiPMs are arrays of SPADs separated by guard rings and other structures, such as trenches to suppress optical cross-talk. Each SiPM SPAD is a reversely biased p-n junction, operated above breakdown. In this configuration, a photo-generated carrier entering the depletion layer may trigger an avalanche. An unfortunate byproduct of the avalanche generation process is the emission of secondary photons, which in some works on SiPM characterisation are referred to as \textit{cross-talk} photons. Secondary photons can be correlated with several factors: electric field, impurity concentrations, doping, geometry, etc.. and are responsible for at least three processes: (i) internal cross-talk (ii) external cross-talk and (iii) optically-induced afterpulsing. With internal cross-talk, we refer to photons that subsequently trigger avalanches in neighbouring SPADs of the same SiPM without escaping from the SiPM itself. With external cross-talk we instead refer to photons that escape from the surface of one SiPM SPAD and potentially: (i) can be reflected back by the SiPM surface coating triggering avalanches in neighbouring SPADs of the same SiPM, (ii) can be transmitted throw the SiPM surface coating leaving the SiPM itself. Finally with optically-induced afterpulsing we refer to secondary photons that trigger avalanches in the same SPAD that originated the primary secondary photon emission during the SPAD recharging time.  This publication focused on the SiPM secondary photon emission outside the SiPM that can be potentially problematic for large surface area, SiPM-based detectors since SiPMs can trigger other SiPMs in their vicinity. For this reason, it is of primary importance to study the SiPM secondary photon emission in order to quantify the systematic effect that this mechanism can produce in the overall detector performance. For this publication we focused on two SiPMs considered as candidates photo-sensor for the nEXO experiment: one FBK~VUV-HD3, and one HPK~VUV4~MPPC. Spectral measurements of their light emission were taken with varying over-voltage in the wavelength range of 450--1020\,nm. At an over voltage of $12.1\pm1.0$~V we measure for the FBK VUV-HD3 a secondary photon yield of  $(4.04\pm0.02)\times 10^{-6}$ $\gamma/e^{-}$. Additionally, the light emitted by the FBK VUV-HD3 shows an interference pattern compatible with thin-film interference, and it presents a low amount of hotspots randomly distributed over the SiPM surface area. For the HPK VUV4 MPCC at an over voltage of $10.7\pm 1.0$~V we measure instead a secondary photon yield of  $(8.71\pm0.04)\times 10^{-6}$ $\gamma/e^{-}$.  Differently from the FBK VUV-HD3, the light emitted by the HPK VUV4 doesn't show an interference pattern, probably due to a thinner surface coating, but it presents a large amounts of hotspots that tend to cluster on one of the corners of the device. Note as the photon yield reported in this paper may be limited if compared with the one reported in previous studies due to the measurement wavelength range that is only up to 1020~nm.
\section{Acknowledgments}\label{sec:acknowledgments}

This work has been supported by: NSERC (Natural Sciences and Engineering Research Council), CFI (Canada Foundation for Innovation) and McDonald Institute in Canada. The authors would like to thank: James Boone and Farid Jalali from Olympus Canada who helped commissioning the IX83 microscope; Nathaniel Kajumba from Delta Photonics Canada who helped commissioning the Princeton Instrument spectrometer system. Additionally, the author would like to thank the nEXO collaboration for their helpful feedback on the manuscript.

\bibliographystyle{unsrt}
\bibliography{LEIM_paper_references}

\begin{thebibliography}{10}

\bibitem{Baudis2018}
L.~Baudis and et~al.
\newblock {Characterisation of Silicon Photomultipliers for liquid xenon
  detectors}.
\newblock {\em Journal of Instrumentation}, 13(10):P10022, 2018.

\bibitem{8490731}
A.~Jamil and et~al.
\newblock Vuv-sensitive silicon photomultipliers for xenon scintillation light
  detection in nexo.
\newblock {\em IEEE Transactions on Nuclear Science}, 65(11):2823--2833, 2018.

\bibitem{CAPASSO2020164478}
M.~Capasso and et~al.
\newblock Fbk vuv-sensitive silicon photomultipliers for cryogenic
  temperatures.
\newblock {\em Nuclear Instruments and Methods in Physics Research Section A},
  982:164478, 2020.

\bibitem{FALCONE2021164648}
A.~Falcone and et~al.
\newblock Cryogenic sipm arrays for the dune photon detection system.
\newblock {\em Nuclear Instruments and Methods in Physics Research Section A},
  985:164648, 2021.

\bibitem{Carnesecchi_2020}
F.~Carnesecchi.
\newblock Light detection in {DarkSide}-20k.
\newblock {\em Journal of Instrumentation}, 15(03):C03038--C03038, 2020.

\bibitem{aalseth2018darkside}
C.~E. Aalseth and et~al.
\newblock Darkside-20k: A 20 tonne two-phase lar tpc for direct dark matter
  detection at lngs.
\newblock {\em The European Physical Journal Plus}, 133(3):131, 2018.

\bibitem{GALLINA2019371}
G.~Gallina and et~al.
\newblock Characterization of the hamamatsu vuv4 mppcs for nexo.
\newblock {\em Nuclear Instruments and Methods in Physics Research Section A},
  940:371--379, 2019.

\bibitem{Acerbi_gain}
F.~Acerbi and et~al.
\newblock Nuv silicon photomultipliers with high detection efficiency and
  reduced delayed correlated-noise.
\newblock {\em IEEE Transactions on Nuclear Science}, 62(3):1318--1325, 2015.

\bibitem{Piemonte_gain}
C.~Piemonte and et~al.
\newblock Performance of nuv-hd silicon photomultiplier technology.
\newblock {\em IEEE Transactions on Electron Devices}, 63(3):1111--1116, 2016.

\bibitem{PhysRev.100.700}
Roger Newman.
\newblock Visible light from a silicon $p\ensuremath{-}n$ junction.
\newblock {\em Phys. Rev.}, 100:700--703, 1955.

\bibitem{SiPMct}
J.~Rosado and et~al.
\newblock {Modeling crosstalk and afterpulsing in silicon photomultipliers}.
\newblock {\em Nuclear Instruments and Methods in Physics Research Section A},
  787:153--156, 2015.

\bibitem{NepomukOtte2009}
A.~{Nepomuk Otte}.
\newblock {On the efficiency of photon emission during electrical breakdown in
  silicon}.
\newblock {\em Nuclear Instruments and Methods in Physics Research Section A},
  610(1):105--109, 2009.

\bibitem{zhang2019light}
Z.~Zhang and et~al.
\newblock Light emission from si avalanche mode leds as a function of e field
  control, impurity scattering, and carrier density balancing.
\newblock In {\em Fifth Conference on Sensors, MEMS, and Electro-Optic
  Systems}, volume 11043, page 1104307. International Society for Optics and
  Photonics, 2019.

\bibitem{LinaLiu2020}
L.~Liu and et~al.
\newblock {2D microspatial distribution uniformity of photon detection
  efficiency and crosstalk probability of multi-pixel photon counters}.
\newblock {\em Quantum Electronics}, 50(2):197--200, 2020.

\bibitem{akil1999multimechanism}
N.~Akil and et~al.
\newblock A multimechanism model for photon generation by silicon junctions in
  avalanche breakdown.
\newblock {\em IEEE Transactions on Electron Devices}, 46(5):1022--1028, 1999.

\bibitem{gautam1988photon}
DK. Gautam and et~al.
\newblock Photon emission from reverse-biased silicon pn junctions.
\newblock {\em Solid-state electronics}, 31(2):219--222, 1988.

\bibitem{Lacaita1993}
A.L. Lacaita and et~al.
\newblock {On the bremsstrahlung origin of hot-carrier-induced photons in
  silicon devices}.
\newblock {\em IEEE Transactions on Electron Devices}, 40(3):577--582, 1993.

\bibitem{gundacker2020silicon}
S.~Gundacker and A.~Heering.
\newblock The silicon photomultiplier: fundamentals and applications of a
  modern solid-state photon detector.
\newblock {\em Physics in Medicine \& Biology}, 65(17):17TR01, 2020.

\bibitem{Boone2017}
K.~Boone and et~al.
\newblock {Delayed avalanches in Multi-Pixel Photon Counters}.
\newblock {\em Journal of Instrumentation}, 12(07):P07026--P07026, 2017.

\bibitem{gallina2021development}
Giacomo Gallina.
\newblock {\em Development of a single vacuum ultra-violet photon-sensing
  solution for nEXO}.
\newblock PhD thesis, University of British Columbia, 2021.

\bibitem{Otte2016}
A.~N. Otte and et~al.
\newblock {Characterization of three high efficiency and blue sensitive silicon
  photomultipliers}.
\newblock {\em Nuclear Instruments and Methods in Physics Research Section A:
  Accelerators, Spectrometers, Detectors and Associated Equipment},
  846:106--125, 2017.

\bibitem{Jamil2018}
A.~Jamil and et~al.
\newblock {VUV-Sensitive Silicon Photomultipliers for Xenon Scintillation Light
  Detection in nEXO}.
\newblock {\em IEEE Transactions on Nuclear Science}, 65(11):2823--2833, 2018.

\bibitem{Acerbi2017}
F.~Acerbi and et~al.
\newblock {Cryogenic Characterization of FBK HD Near-UV Sensitive SiPMs}.
\newblock {\em IEEE Transactions on Electron Devices}, 64(2):521--526, feb
  2017.

\bibitem{engelmann2018spatially}
E.~Engelmann and et~al.
\newblock Spatially resolved dark count rate of sipms.
\newblock {\em The European Physical Journal C}, 78(11):1--8, 2018.

\bibitem{Nagai2018}
A.~Nagai and et~al.
\newblock {Characterisation of a large area silicon photomultiplier}.
\newblock oct 2018.

\bibitem{mirzoyan2009light}
R~Mirzoyan and et~al.
\newblock Light emission in si avalanches.
\newblock {\em Nuclear Instruments and Methods in Physics Research Section A},
  610(1):98--100, 2009.

\bibitem{kittel1996introduction}
Charles Kittel.
\newblock {\em Introduction to solid state physics}, volume~6.
\newblock Wiley New York, 1986.

\bibitem{mcclure2010intellical}
Jason McClure and Ed~Gooding.
\newblock Intellical: A novel method for calibration of imaging spectrographs.
\newblock In {\em AIP Conference Proceedings}, volume 1267, pages 806--807.
  American Institute of Physics, 2010.

\bibitem{Jackson2002}
J.~C. Jackson and et~al.
\newblock {Comparing leakage currents and dark count rates in Geiger-mode
  avalanche photodiodes}.
\newblock {\em Applied Physics Letters}, 80(22):4100--4102, 2002.

\bibitem{PIbook}
Princeton Instrument.
\newblock {PyLoN System User Manual}.
\newblock
  \url{https://www.princetoninstruments.com/wp-content/uploads/2020/04/PyLoN-System-Manual-Issue-5-4411-0136.pdf},
  2007.

\bibitem{VUVHD3_reflectance}
P.~Lv and et~al.
\newblock Reflectance of silicon photomultipliers at vacuum ultraviolet
  wavelengths.
\newblock {\em IEEE Transactions on Nuclear Science}, 67(12):2501--2510, 2020.

\bibitem{griffiths1962introduction}
David~J Griffiths.
\newblock {\em Introduction to electrodynamics}.
\newblock Prentice Hall New Jersey, 1962.

\bibitem{green_self-consistent_2008}
A.~G. Martin.
\newblock Self-consistent optical parameters of intrinsic silicon at 300~k
  including temperature coefficients.
\newblock {\em Solar Energy Materials and Solar Cells},
  92:1305{\textendash}1310, 2008.

\bibitem{8818357}
G.~Gallina and et~al.
\newblock Characterization of sipm avalanche triggering probabilities.
\newblock {\em IEEE Transactions on Electron Devices}, 66(10):4228--4234, 2019.

\bibitem{aspnes1983dielectric}
David~E Aspnes and AA~Studna.
\newblock Dielectric functions and optical parameters of si, ge, gap, gaas,
  gasb, inp, inas, and insb from 1.5 to 6.0 ev.
\newblock {\em Physical review B}, 27(2):985, 1983.

\bibitem{rodriguez2016self}
Luis~V. Rodr\'{i}guez-de Marcos and et~al.
\newblock Self-consistent optical constants of sio2 and ta2o5 films.
\newblock {\em Opt. Mater. Express}, 6(11):3622--3637, 2016.

\bibitem{wolff1960theory}
PA~Wolff.
\newblock Theory of optical radiation from breakdown avalanches in germanium.
\newblock {\em Journal of Physics and Chemistry of Solids}, 16(3-4):184--190,
  1960.

\bibitem{haecker1974infrared}
W~Haecker.
\newblock Infrared radiation from breakdown plasmas in si, gasb, and ge:
  Evidence for direct free hole radiation.
\newblock {\em physica status solidi (a)}, 25(1):301--310, 1974.

\bibitem{figielski1962proc}
T~Figielski and A~Torun.
\newblock Proc. int. conf. on semiconductor physics.
\newblock 1962.

\bibitem{Piemonte2016}
C.~Piemonte and et~al.
\newblock {Performance of NUV-HD Silicon Photomultiplier Technology}.
\newblock {\em IEEE Transactions on Electron Devices}, 63(3):1111--1116, 2016.

\end{thebibliography}
\end{document}